\documentclass[12pt]{iopart}
\usepackage{graphicx}
\usepackage{color}
\usepackage{comment}

\expandafter\let\csname equation*\endcsname\relax

\expandafter\let\csname endequation*\endcsname\relax

\usepackage{amsmath}
\usepackage{amssymb}
\usepackage{float}

\def\bk{{\bold{k}}}
\def\bb{{\bold{b}}}
\def\bs{{\bold{s}}}
\def\bh{{\bold{h}}}

\def\bd{{\bold{d}}}

\def\bq{{\bold{q}}}
\def\bg{{\bold{g}}}

\def\m1{{^{-1}}}

\begin{document}

\title{Nonunitary Superconductivity in Complex Quantum Materials}

\author{Aline Ramires}

\address{Paul Scherrer Institute, CH-5232 Villigen PSI, Switzerland}

\ead{aline.ramires@psi.ch}

\begin{abstract}
We present a comprehensive discussion on nonunitary superconductivity in complex quantum materials. Starting with a brief review of the notion of nonunitary superconductivity, we discuss its spectral signatures in simple models with only the spin as an internal degree of freedom. In complex materials with multiple internal degrees of freedom, there are many more possibilities for the development of nonunitary order parameters. We provide examples focusing on d-electron systems with two orbitals, applicable to a variety of materials. We discuss the consequences for the superconducting spectra, highlighting that gap openings of band crossings at finite energies can be attributed to a nonunitary order parameter if this is associated with a finite superconducting fitness matrix $\hat{F}_C(\bk)$.  We speculate that nonunitary superconductivity in complex quantum materials is in fact very common and can be associated with multiple cases of time-reversal symmetry breaking superconductors.
\end{abstract}

\maketitle


\section{Introduction}

Superconducting complex quantum materials have recently attracted a lot of attention given their unusual phenomenology. Some heavy fermion materials, such as CeRh$_2$As$_2$ \cite{Khim:2021} and UTe$_2$ \cite{Ran:2019} host multiple superconducting phases, Sr$_2$RuO$_4$ has been puzzling the condensed matter community with experimental results that are hard to reconcile \cite{Mac:2017}, doped topological insulators in the family of Bi$_2$Se$_3$ display unusual robustness in presence of impurities despite the unconventional nature of the order parameter \cite{Andersen:2020}, and transition metal dichacolgenides have upper critical fields much larger than the one set by the Pauli limit \cite{Lu:2015}. From a theoretical perspective, what is common among these families of materials is the presence of multiple internal degrees of freedom (DOFs), such as orbitals, sublattices, or valleys, in the respective effective models for the normal state electronic structure. 

The presence of internal DOFs in the effective description of the electronic structure of these materials opens new possibilities for the internal structures of the superconducting order parameter. For simple superconductors, with only the spin as an internal DOF, the only possibilities for pairing are even-parity spin-singlet, or odd-parity spin-triplet states. In presence of one  extra internal DOF, one can now have pairs that are symmetric or anti-symmetric with respect to this new DOF. For example, even-parity spin-triplet orbital-antisymmetric states are possible, among others. With a single extra internal DOF which can acquire two values (such as an orbital DOF that can be associated with $d_{xz}$ or $d_{yz}$) the number of order parameter realizations jumps to sixteen, in contrast to four in case the spin is the only internal DOF (one singlet and three triplet states).

This new richness in order parameter space allows for multiple realizations of the order parameter to fall within the same irreducible representation. In complex superconductors, order parameters are generally a linear superposition of different order parameter realizations within the same symmetry classification. As we discuss in detail below, this linear superposition leads to a non-unitary order parameter. Given a superconducting state characterized by the order parameter matrix $\hat{\Delta}(\bk)$, if the gauge invariant combination $\hat{\Delta}(\bk)\hat{\Delta}^\dagger(\bk)$ is proportional to the identity matrix the superconducting state is called unitary, if it is not simply proportional to the identity, the order parameter is generally called nonunitary. Most superconducting states discussed in the literature are unitary. Recently, nonunitary superconductors started to attract special attention given its association with Bogoliubov Fermi surfaces \cite{Brydon:2018,Agterberg:2017}, anapole superconductivity \cite{Kanasugi:2021}, and topolgical classification in terms of q-helicity \cite{Hatsugai:2004}.

Nonunitary states were first discussed in the context of the A$_1$ phase in superfluid He-3 in the presence of an applied magnetic field \cite{Leggett:1975,Wheatley:1975}. The first mention of nonunitary superconducting states was made in the context of UPt$_3$ \cite{Ohmi:1993,Machida:1998,Tou:1998}. These ideas have been recently revisited considering the nonsymmorphic structure of this material \cite{Yanase:2016}. Recently, nonunitary superconductity has also been proposed for the time-reversal symmetry breaking superconductors LaNiC$_2$ \cite{Csire:2018a,Hillier:2009,Quintanilla:2010,Chen:2013} and LaNiGa$_2$ \cite{Hillier:2012,Weng:2016,Ghosh:2020},
the former noncentrosymmetric, and the later centrosymmetric. The newly reported multiple superconducting phases in UTe$_2$ \cite{Braithwaite:2019} could also be associated with nonunitary order parameters  based on the magnetic space group classification of the order parameter \cite{Yarzhemsky:2020}, or on triplet pairing on the border of magnetism \cite{Nevidomskyy:2020}.

Theoretical works in the context of nonunitary superconductivity have explored disorder-induced mixed-parity superconductivity in noncentrosymmetric monolayer transition-metal dichalcogenides \cite{Mockli:2018}, and field-induced mixed-parity in locally noncentrosymmetric materials \cite{Yoshida:2014}. Experimental observables such as the magnetoelectric Andreev effect  \cite{Tkachov:2017} and signatures in the conductance spectra in ferromagnetic metal/nonunitary superconductor junctions \cite{Linder:2007} were proposed. More recently, the opening of gaps away from the Fermi surface in Dirac materials was suggested as a signature of multi-orbital nonunitary superconductivity \cite{Lado:2019}.

In this paper, we propose a comprehensive discussion on nonunitary superconductivity in complex quantum materials. We start introducing the concept of superconducting fitness, which is going to be key in the later discussion of spectral signatures for both simple and complex scenarios. We briefly review the notion of nonunitary order parameters in simple superconductors, emphasizing that the development of nonunitary superconductivity is directly associated with symmetry-breaking order parameters. Without delving into the details of the pairing mechanism, we are going to discuss the spectral signatures of these order parameters. We highlight that in the context of simple superconductors, gap openings at high energy particle-hole band crossings are primarily associated with a non-zero superconducting fitness matrix, not with the nonunitary character of the order parameter.  In the context of complex materials, with multiple internal DOFs, nonunitary order parameters can develop without extra symmetry breaking. We highlight some examples of nonunitary pairing in complex materials focusing on d-electron systems with two orbitals. We discuss the consequences for the superconducting spectrum in these more complex scenarios based on our refined understanding of nonunitary superconductivity in simple systems. We conclude providing a discussion that connects with multiple recent works, and we speculate on the connections between nonunitary and time-reversal symmetry breaking superconductors in complex materials.

\section{The concept of superconducting fitness}
\label{sec:fitness}

Before staring the discussion on nonunitary superconductivity, it is useful to introduce the concept of \emph{superconducting fitness} \cite{Ramires:2016,Ramires:2018}, as this is going to be a key ingredient in the analysis that follows. The superconducting fitness framework has been a useful theoretical tool allowing for the understanding of the stability and nodal structure of unexpected superconducting states \cite{Ramires:2019,Suh:2020,Ramires:2021,Mockli:2021}, the robustness of unconventional superconducting states in the presence of impurities \cite{Andersen:2020, Zinkl:2022}, and the unusual behaviour of complex superconductors under external symmetry breaking fields, such as strain \cite{Beck:2021}. This seems to be one of the most appropriate frameworks for the description of superconductivity in complex quantum materials, allowing for the understanding of apparently contradicting responses of complex superconductors.

We start introducing the effective  Bogoliubov-de Gennes (BdG) Hamiltonian:
\begin{eqnarray}
H_{BdG} = \sum_\bk \Psi_{\bk}^\dagger \hat{\mathcal{H}}_{BdG} (\bk)\Psi_{\bk},
\end{eqnarray}
where
\begin{eqnarray}\label{Eq:BdG}
\hat{\mathcal{H}}_{BdG}(\bk) =  \begin{pmatrix} 
\hat{H}_0(\bk) & \hat{\Delta}(\bk)\\
\hat{\Delta}^\dagger(\bk) & - \hat{H}_0^*(-\bk)
 \end{pmatrix},
\end{eqnarray}
and $ \Psi_{\bk}^\dagger  = (\Phi_\bk^\dagger, \Phi_\bk^T)$ is a Nambu spinor, where $\Phi_\bk^\dagger= (c_{\bk 1\uparrow}^\dagger, c_{\bk 1\downarrow}^\dagger, c_{\bk 2\uparrow}^\dagger, c_{\bk 2\downarrow}^\dagger,..., c_{\bk n\uparrow}^\dagger, c_{\bk n\downarrow}^\dagger)$, and $c_{\bk \alpha \sigma}^\dagger$ ($c_{\bk \alpha \sigma}$)  creates (annihilates) an electron with momentum $\bk$ and spin $\sigma$ in the internal DOF $\alpha$, which can be associated with orbitals, sublattice, or valley structures. $\hat{H}_0(\bk)$ corresponds to the normal state Hamiltonian, and $\hat{\Delta}(\bk)$ encodes the order parameter, both are $2n \times 2n$ matrixes in case the parameter $\alpha$ can acquire $n$ different values.

\subsection{Superconducting Fitness and inter-band pairing}

Superconductivity is usually discussed as an electronic instability out of a metallic state characterized by one or more bands. Even in the case of a single internal DOF, such as models with spinfull electrons on a single orbital, we can find multiple bands in presence of external symmetry breaking fields. In case of multiple internal DOFs, we naturally find multiple electronic bands, which can be doubly degenerate in presence of time-reversal and inversion symmetries. If pairing happens between electrons in the same band (intra-band pairing), superconductivity is stablished for an arbitrarily small attractive interaction through the Cooper instability with the formation of superconducting pairs of electrons with total zero momentum. If pairing happens between electrons in different bands (inter-band pairing),  the superconducting state is not as stable since a finite attractive interaction is necessary. Based on these ideas, below we present an heuristic discussion and introduction of the concept of superconducting fitness \cite{Ramires:2016}.

In presence of external symmetry breaking fields or multiple orbitals or sublattices, the normal state Hamiltonian $\hat{H}_0(\bk)$ is generally not diagonal in the microscopic basis (based on the orbital and sublattice DOFs). As the Hamiltonian is an Hermitian matrix, there is always a unitary transformation $\hat{U}(\bk)$ which diagonalizes it, or rotates it to the band basis: $\hat{H}^B_0(\bk)=\hat{U}(\bk) \hat{H}_0(\bk) \hat{U}^\dagger (\bk)$ (the superscript $B$ stands for the band, or diagonal, basis). The gap matrix, by connecting particle-particle spaces, transforms in a slightly different manner: $\hat{\Delta}^B(\bk) =\hat{U}(\bk) \hat{\Delta}(\bk) \hat{U}^T(-\bk)$. In case of pure intra-band pairing $\Delta^B(\bk)$ is block diagonal. A non block-diagonal gap matrix is an indication of inter-band pairing. 

To get some intuition on the origin of the concept of superconducting fitness, we consider the minimal multi-orbital problem consisting of two orbitals. In presence of time reversal and inversion symmetries,  $\hat{H}_0^B(\bk)$ has doubly degenerate states with energy $\epsilon_a$, where $a$ is the band label, and therefore has a structure with $2 \times 2$ blocks proportional to the identity $\hat{\sigma}_0$. Note that for an arbitrary gap matrix with intra- $\hat{\Delta}_{a}$ and inter- $\hat{\Delta}_{ab}$ band components, we have, omitting the momentum dependence:
\begin{eqnarray}\label{MFH2}
\hat{H}^B_0 =\begin{pmatrix}
\epsilon_1 \hat{\sigma}_0 & 0\\
0& \epsilon_2 \hat{\sigma}_0
\end{pmatrix},\hspace{0.5cm}
\hat{\Delta}^B  &=& \begin{pmatrix}
\hat{\Delta}_1 & \hat{\Delta}_{12}\\
\hat{\Delta}_{21}& \hat{\Delta}_2
\end{pmatrix}.
\end{eqnarray}
Note that these matrices do not commute for finite inter-band pairing, unless the artificial condition $\epsilon_1=\epsilon_2$ is satisfied. On the other hand, in case $\hat{\Delta}_{ab}(\bk)=0$, $\hat{\Delta}^B(\bk)$ is block diagonal and commutes with the bare Hamiltonian $\hat{H}^B_0(\bk)$ in the band basis.

We can now look at the condition for absence of inter-band pairing from the microscopic basis perspective. Using the unitary transformation introduced above and the fact that $\hat{H}^B_0(\bk)$ and $\hat{\Delta}^{B}(\bk)$ commute in case of pure intra-band pairing, we can write:
\begin{eqnarray}
\hat{H}_0(\bk) \hat{\Delta}(\bk) &=& \hat{U}^\dagger(\bk) \hat{H}^B_0(\bk) \hat{U}(\bk)\hat{U}^\dagger(\bk)  \hat{\Delta}^{B}(\bk) \hat{U}^*(-\bk) ,\\
&=& \hat{U}^\dagger(\bk)  \hat{H}^B_0(\bk)  \hat{\Delta}^{B}(\bk) \hat{U}^*(-\bk)\nonumber\\
&=& \hat{U}^\dagger(\bk)   \hat{\Delta}^{B}(\bk)  \hat{H}^B_0(\bk) \hat{U}^*(-\bk)\nonumber\\
&=& \hat{U}^\dagger(\bk)   \hat{\Delta}^{B}(\bk)\hat{U}^*(-\bk) \hat{U}^T(-\bk)   \hat{H}^B_0(\bk) \hat{U}^*(-\bk).\nonumber
\end{eqnarray}
We can identify the first three factors in the last line with $\hat{\Delta}(\bk)$. For the last three factors, we use inversion symmetry (already assumed above to guarantee the double degeneracy of the bands) and the fact that the eigenvalues of $\hat{H}_0^B(\bk)$ are real to write $\hat{H}^*_0(-\bk)=  \hat{U}^T(-\bk) \hat{H}^B_0(\bk) \hat{U}^*(-\bk)$, so we have:
\begin{eqnarray}
\hat{H}_0(\bk) \hat{\Delta}(\bk) - \hat{\Delta}(\bk)\hat{H}^*_0(-\bk) = 0.
\end{eqnarray}
If $\hat{H}_0(\bk)$ and $\hat{\Delta}(\bk)$ satisfy this condition, the system develops only intra-band pairing and consequently has a robust superconducting instability. In case the identity above is not satisfied, we have a measure of the incompatibility between the superconducting state and the normal state, associated with the presence of inter-band pairing, which we label as the superconducting fitness matrix $\hat{F}_C(\bk)$:
\begin{eqnarray}\label{eq:FC}
\hat{F}_C(\bk)= \hat{H}_0(\bk) \hat{\Delta}(\bk) - \hat{\Delta}(\bk)\hat{H}^*_0(-\bk).
\end{eqnarray}

\subsection{Superconducting Fitness and the critical temperature}

A finite superconducting fitness matrix $\hat{F}_C(\bk)$ has been shown to be directly associated with the suppression of the superconducting critical temperature from its maximum possible value in single band superconductors  \cite{Ramires:2016}. For models with two orbitals or sublattices,  the superconducting critical temperature can be explicitly written as \cite{Ramires:2018}:
\begin{eqnarray}\label{eq:Tc}
k_B T_C &=&  \frac{4 e^\gamma}{\pi}\frac{\omega_C}{2 } e^{-1/2|v| \alpha} e^{-\delta /\alpha},
\end{eqnarray}
where $\gamma$ is the Euler constant, $\omega_C$ is an energy cutoff, $|v|$ is the magnitude of the effective attractive interaction on the superconducting channel of interest. In addition, 
\begin{eqnarray}
\delta = \frac{\omega_C^2}{32}\sum_a N_a(0)  \left\langle \frac{Tr[|\hat{F}_C(\bk)|^2]}{q(\bk_{Fa})^2} \right\rangle_{\!\!FS_a},
\end{eqnarray}
and
\begin{eqnarray}
\alpha = \frac{1}{16} \sum_a N_a(0) \langle Tr[|\hat{F}_A(\bk)|^2]\rangle_{FS_a},
\end{eqnarray}
where the sum over the index $a$ runs over the bands with density of states at the Fermi surface $N_a(0)$, and $\langle...\rangle_{FS_a}$ denotes the average over the respective Fermi surface. Here  $q(\bk)= \epsilon_a(\bk)-\epsilon_b(\bk)$ is the energy difference between the two bands, which we assume are well separated, $q(\bk)>> \omega_C$. The matrix inside the trace for the parameter $\alpha$ is the anti-commutator counterpart of the first superconducting fitness matrix defined above:
\begin{eqnarray}\label{eq:FA}
\hat{F}_A(\bk)= \hat{H}_0(\bk) \hat{\Delta}(\bk) + \hat{\Delta}(\bk)\hat{H}^*_0(-\bk).
\end{eqnarray}

From the closed form equation for the critical temperature, Eq. \ref{eq:Tc}, it is clear that the largest the parameter $\alpha$ [so the larger $\hat{F}_A(\bk)$], the largest the critical temperature, while a finite $\delta$ [or a finite $\hat{F}_C(\bk)$] suppresses the critical temperature from its potentially maximum value.

\subsection{Superconducting Fitness and odd frequency correlations}

Recently, the superconducting fitness measure $\hat{F}_C(\bk)$ was also associated with the presence of odd-frequency superconducting correlations \cite{Triola:2020}. This can be shown by manipulating the BdG Green's function, defined as:
\begin{eqnarray}
\hat{G}_{BdG}(\bk, i\omega_m) = \left[i\omega_m\hat{I}_{2n} -\hat{H}_{BdG}(\bk)\right]^{-1},
\end{eqnarray}
where $\hat{I}_{2n}$ is the $2n \times 2n$ identity matrix, and $\omega_m$ is a Matsubara frequency. This Green's function can be broken down in particle and hole spaces, such that we can write:
\begin{eqnarray}
\hspace{-1cm}
\begin{pmatrix}
\hat{G}_{e}(\bk, i\omega_m) & \hat{F}(\bk,i\omega_m)\\
\hat{F}^\dagger(\bk,i\omega_m) & \hat{G}_{h}(\bk, i\omega_m) 
\end{pmatrix}
.
\begin{pmatrix}
\hat{I}_n i\omega_m - \hat{H}_0(\bk) & -\hat{\Delta}(\bk) \\
 -\hat{\Delta}^\dagger(\bk) & \hat{I}_n i\omega_m + \hat{H}_0^*(-\bk) 
\end{pmatrix} 
=\hat{I}_{2n}.
\end{eqnarray}

From the equation above we can extract two identities:
\begin{eqnarray}
\hat{G}_{e}(\bk, i\omega_m) (\hat{I}_n i\omega_m - \hat{H}_0(\bk) ) -  \hat{F}(\bk,i\omega_m)\hat{\Delta}^\dagger(\bk)  = \hat{I}_n,\\
- \hat{G}_{e}(\bk, i\omega_m) \hat{\Delta}(\bk)  +  \hat{G}_{h}(\bk, i\omega_m)  (\hat{I}_n i\omega_m + \hat{H}_0^*(-\bk) ) = 0.
\end{eqnarray}
Isotaling $\hat{G}_{e}(\bk, i\omega_m)$ from the first equation, substituting it in the second and identifying $\hat{G}_{e}^0(\bk, i\omega_m) = (\hat{I}_n i\omega_m - \hat{H}_0(\bk) )^{-1}$ and $\hat{G}_{h}^0(\bk, i\omega_m) = (\hat{I}_n i\omega_m + \hat{H}_0^*(-\bk) ) ^{-1}$ as the normal state particle and hole Green's functions, we can write:
\begin{eqnarray}
\hat{F}(\bk,i\omega_m) = \hat{F}^0(\bk,i\omega_m) [\hat{I}_n -  \hat{\Delta}^\dagger(\bk)\hat{F}^0(\bk,i\omega_m)]^{-1},
\end{eqnarray}
where we identified $\hat{F}^0(\bk,i\omega_m) =\hat{G}_{e}^0(\bk, i\omega_m) \hat{\Delta}(\bk) \hat{G}_{h}^0(\bk, i\omega_m) $ as the first order contribution in $\hat{\Delta}(\bk)$ to the anomalous correlation function $\hat{F}(\bk,i\omega_m)$.

Note that all the frequency dependence in $\hat{F}(\bk,i\omega_m)$ appears through $\hat{F}^0(\bk,i\omega_m)$, such that if the last is odd in frequency, the first should also be. This simplifies the discussion, as we can consider the simpler form of $\hat{F}^0(\bk,i\omega_m)$ to extract the condition for the presence of odd frequency anomalous correlations. Writing $\hat{F}^0(\bk,i\omega_m)$ explicity:
\begin{eqnarray}
\hat{F}^0(\bk,i\omega_m) = (\hat{I}_n i\omega_m - \hat{H}_0(\bk) )^{-1} \hat{\Delta}(\bk)(\hat{I}_n i\omega_m + \hat{H}_0^*(-\bk) ) ^{-1},
\end{eqnarray}
we can make the first and last factors even in frequency by multiplying them by a convenient identity factor. After some manipulation we find:
\begin{eqnarray}
\hspace{-1cm}
\hat{F}^0(\bk,i\omega_m) &=& \{\hat{I}_n (i\omega_m)^2 - [\hat{H}_0(\bk)]^2 \}^{-1} \\ \nonumber &&\times 
[( i\omega_m)^2\hat{\Delta}(\bk) - \hat{H}_0(\bk)\hat{\Delta}(\bk)  \hat{H}_0^*(-\bk) + i\omega_m \hat{F}_C(\bk)]\\ \nonumber
&&\times\{\hat{I}_n( i\omega_m )^2 - [\hat{H}_0^*(-\bk)]^2 \}^{-1}.
\end{eqnarray}

From the last line, it is evident that the presence of odd frequency correlations is directly associated with a nonzero superconducting fitness matrix $\hat{F}_C(\bk)$:
\begin{eqnarray}
\hspace{-1cm}
\hat{F}^{0}_{odd}(\bk,i\omega_m) &=& \hat{F}^0(\bk,i\omega_m) - \hat{F}^0(\bk,-i\omega_m) \\ \nonumber
&=& i\omega_m \{\hat{I}_n (i\omega_m)^2 - [\hat{H}_0(\bk)]^2 \}^{-1}
 \hat{F}_C(\bk)\{\hat{I}_n( i\omega_m )^2 - [\hat{H}_0^*(-\bk)]^2 \}^{-1}.
\end{eqnarray}

\section{Superconductivity in simple metals}\label{sec:simple}

The normal state Hamiltonian, $\hat{H}_0(\bk)$, has the general form:
\begin{eqnarray}
\hat{H}_0(\bk)  = \epsilon(\bk) \hat{\sigma}_0 + \bs (\bk) \cdot \hat{\boldsymbol{\sigma}},
\end{eqnarray}
where $\epsilon(\bk)$ describes a doubly degenerate electronic dispersion in presence of time-reversal symmetry (TRS) and inversion symmetry (IS). The three-dimensional vector $\bs(\bk)$ introduces time-reversal symmetry breaking (TRSB) as an external magnetic field, if $\bs(\bk) = - \bb$, or inversion symmetry breaking (ISB) as odd spin-orbit coupling (SOC), if $\bs(\bk) =  \bg(\bk) = - \bg(-\bk)$. Here $\hat{\sigma}_0$ is the two-dimensional identity matrix and $\hat{\boldsymbol{\sigma}} = (\hat{\sigma}_x,\hat{\sigma}_y,\hat{\sigma}_z)$ is a three-dimensional vector of Pauli matrices.

The superconducting order parameter, $\hat{\Delta}(\bk)$, can be written as
\begin{eqnarray}\label{eq:gap}
\hat{\Delta}(\bk) = [d_0(\bk)+ \bd(\bk) \cdot \hat{\boldsymbol{\sigma}}](i\hat{\sigma}_2).
\end{eqnarray}
Following fermionic antisymmetry, the order parameter matrix should satisfy $\hat{\Delta}(\bk) = - \hat{\Delta}^T(-\bk)$. As a consequence, $d_0(\bk)$ is an even function of momenta, while $\bd(\bk)$ is an odd function of momenta.   Time-reversal symmetry is implemented as $\hat{\Theta} = (i\hat{\sigma}_2) \mathcal{K}$, where $\mathcal{K}$ stands for complex conjugation, and should be accompanied by the change in momenta $\bk \rightarrow -\bk$. This definition, when applied to the superconducting gap gives 
$\hat{\Delta}_T(\bk) = \Theta \hat{\Delta}(\bk) \Theta^{-1} 
= d_0^*(\bk) (i\sigma_2) + \bd^*(\bk) \cdot \hat{\boldsymbol{\sigma}}(i\sigma_2)$,
such that if we choose the function $d_0(\bk)$ and all components of $\bd(\bk)$ to have the same phase, the order parameter is time-reversal invariant [up to an overall $U(1)$ gauge transformation].

A gauge-invariant composition of the order parameter with itself is 
\begin{eqnarray}
\hat{\Delta}(\bk)\hat{\Delta}^\dagger(\bk) = \Delta^2_{U}(\bk) \hat{\sigma}_0 + \bq_{NU}(\bk)\cdot \hat{\boldsymbol{\sigma}},
\end{eqnarray} 
where 
\begin{eqnarray}
\Delta^2_{U}(\bk) =  \left[|d_0(\bk)|^2 + |\bd(\bk)|^2\right]
\end{eqnarray} 
is the magnitude of the unitary part of the gap, and 
\begin{eqnarray}
\bq_{NU}(\bk) = \left[d_0(\bk) \bd^*(\bk) + d_0^*(\bk) \bd (\bk) + i \bd(\bk) \times \bd^*(\bk) \right],
\end{eqnarray} 
is the vector that characterizes the nonunitary part of the gap. As a definition, the order parameter is unitary if $\bq_{NU}(\bk)  = 0$, and  is nonunitary otherwise. From the explicit form of $\bq_{NU}(\bk)$, we can conclude that the order parameter is nonunitary if at least one of the following conditions are met: i) ISB in the normal state, allowing parity mixing in the superconducting state; ii) TRSB order parameter with $\bd^*(\bk)$ not parallel to $\bd(\bk)$. Note that only in the later case the order parameter spontaneously breaks the mentioned symmetry. In the case of an ISB order parameter, inversion symmetry is already broken in the normal state, so it is not spontaneously broken in the superconducting state. 

A nonunitary order parameter leads to a finite expectation value of $\langle \hat{\Delta}^\dagger(\bk) \hat{\boldsymbol{\sigma}}\hat{\Delta}(\bk)\rangle$, which is associated with a finite spin polarization of the pair at a given $\bk$. Note that the average over the Fermi surface can lead to a zero net spin polarization [as in the case of ISB in the normal state, since $d_0(\bk)$ is even and $\bd(\bk)$ is odd in momenta]. This tells us that only a TRSB superconductor could sustain a finite average spin polarization of the pair. 

In this regard, it is useful to refine the notion of nonunitary order parameters introducing the time-reversal-odd gauge-invariant product \cite{Brydon:2018}:
\begin{eqnarray}
\hat{\Delta}(\bk)\hat{\Delta}^\dagger(\bk)  - \hat{\Delta}_T(\bk)\hat{\Delta}_T^\dagger(\bk) = \bq_{TRO}(\bk) \cdot  \hat{\boldsymbol{\sigma}}
\end{eqnarray} 
where $ \hat{\Delta}_T(\bk)$ is the time-reversed order parameter defined above, and
\begin{eqnarray}
\bq_{TRO}(\bk) = 2i \bd(\bk) \times \bd^*(\bk).
\end{eqnarray} 
If $\bq_{TRO}(\bk) \neq 0$ the TRSB order parameter develops a spin polarization. It has recently been proposed that a finite $\bq_{TRO}(\bk)$, and not a finite $\bq_{NU}(\bk)$, should be taken as the definition of nonunitary pairing \cite{Brydon:2018}.

\section{Signatures of nonunitary order parameters in the single orbital scenario}
\label{sec:signatures}

The energy spectrum in the superconducting state can be evaluated by diagonalizing the BdG Hamiltonian in Eq. \ref{Eq:BdG}. To understand some of the features of the spectrum, it is convenient to take the square of the BdG Hamiltonian, which give us as eigenvalues the squares of the eigenenegies. The explicit form of the square of the BdG Hamiltonian is very suggestive:
\begin{eqnarray}\label{eq:square}
\hat{\mathcal{H}}_{BdG}^2 (\bk)&=&\begin{pmatrix} 
\hat{H}_0(\bk) & \hat{\Delta}(\bk)\\
\hat{\Delta}^\dagger(\bk) & - \hat{H}_0^*(-\bk)
 \end{pmatrix}^2 \\ \nonumber
 &=&\begin{pmatrix} 
[\hat{H}_0(\bk)]^2 + \hat{\Delta}(\bk)\hat{\Delta}^\dagger(\bk)&\hat{F}_C(\bk)\\
\hat{F}^\dagger_C(\bk) & [ \hat{H}_0^*(-\bk) ]^2 + \hat{\Delta}^\dagger(\bk)\hat{\Delta}(\bk)
 \end{pmatrix}.
\end{eqnarray}
where $\hat{F}_C(\bk)$ is the superconducting fitness matrix, defined in Eq. \ref{eq:FC}. A summary of the superconducting fitness matrices for a single internal DOF (spin)  is given in Table \ref{table:fit1}. 

Below we analyze three different senarios: unitary, nonunitary associated with ISB [$\bq_{TRO} (\bk) = 0$] and nonunitary associated with TRSB [$\bq_{TRO} (\bk) \neq 0$] . In each case we highlight the role of the superconducting fitness measure by taking examples with $\hat{F}_C(\bk)=0$ and $\hat{F}_C(\bk)\neq0$.

\begin{center}
\begin{table}[h]
\begin{tabular}{|c|c|c|}
\cline{2-3}
\multicolumn{1}{c|}{}
& \begin{tabular}{c}TRSB\\ $\bs(\bk)=-\mathbf{h}$\end{tabular}& 
\begin{tabular}{c} ISB \\
$\bs(\bk)=\mathbf{g}(\bk)$\end{tabular}\\ 
\cline{1-3}
 \begin{tabular}{c}Singlet\\ $\hat{\Delta}_S(\bk) = d_0(\bk) (i\hat{\sigma}_2)$\end{tabular}
&  Always finite &  Always zero\\
\cline{1-3}
 \begin{tabular}{c}Triplet\\ $\hat{\Delta}_T(\bk) = \bd(\bk) \cdot \hat{\boldsymbol{\sigma}}(i\hat{\sigma}_2)$\end{tabular}
 &$ \propto \mathbf{h}\cdot \mathbf{d}(\bk)$ & $ \propto \mathbf{g}(\bk) \times \mathbf{d}(\bk)$\\
\cline{1-3}
\end{tabular}
\caption{Superconducting fitness for singlet and triplet superconducting states under time reversal symmetry breaking (TRSB) or inversion symmetry breaking (ISB).}
\label{table:fit1}
\end{table}
\end{center}

\subsection{Case I: Unitary order parameter}\label{sec:casei}

Let's start with the simpler case with $\hat{F}_C(\bk)=0$, so the square of the BdG Hamiltonian matrix, Eq. \ref{eq:square},  is block diagonal. The upper left block reads: 
\begin{eqnarray}
[\hat{H}_0(\bk)]^2 + \hat{\Delta}(\bk)\hat{\Delta}^\dagger(\bk) &=&  \{[\epsilon(\bk)]^2 + [\mathbf{s}(\bk)]^2 + |\Delta_U(\bk)|^2\} \hat{\sigma}_0 \\ \nonumber &&+ 2 \epsilon(\bk)\mathbf{s}(\bk) \cdot \hat{\boldsymbol{\sigma}}.
\end{eqnarray}

The eigenvectors of the square of the BdG Hamiltonian are then 
\begin{eqnarray}
E^2_\pm &=& [\epsilon(\bk)]^2 + [\mathbf{s}(\bk)]^2 + |\Delta_U(\bk)|^2 \pm |2 \epsilon(\bk)\mathbf{s}(\bk)| \\ \nonumber
&=& [\epsilon(\bk) \pm |\mathbf{s}(\bk)|]^2 + |\Delta_U(\bk)|^2,
\end{eqnarray}
such that the eigenvalues of the original BdG Hamiltonian are:
\begin{eqnarray}
E_{\pm\pm} &=& \pm \sqrt{ [\epsilon(\bk) \pm |\mathbf{s}(\bk)|]^2 + |\Delta_U(\bk)|^2}.
\end{eqnarray}

\begin{figure}[h]
\includegraphics[width=0.8\textwidth]{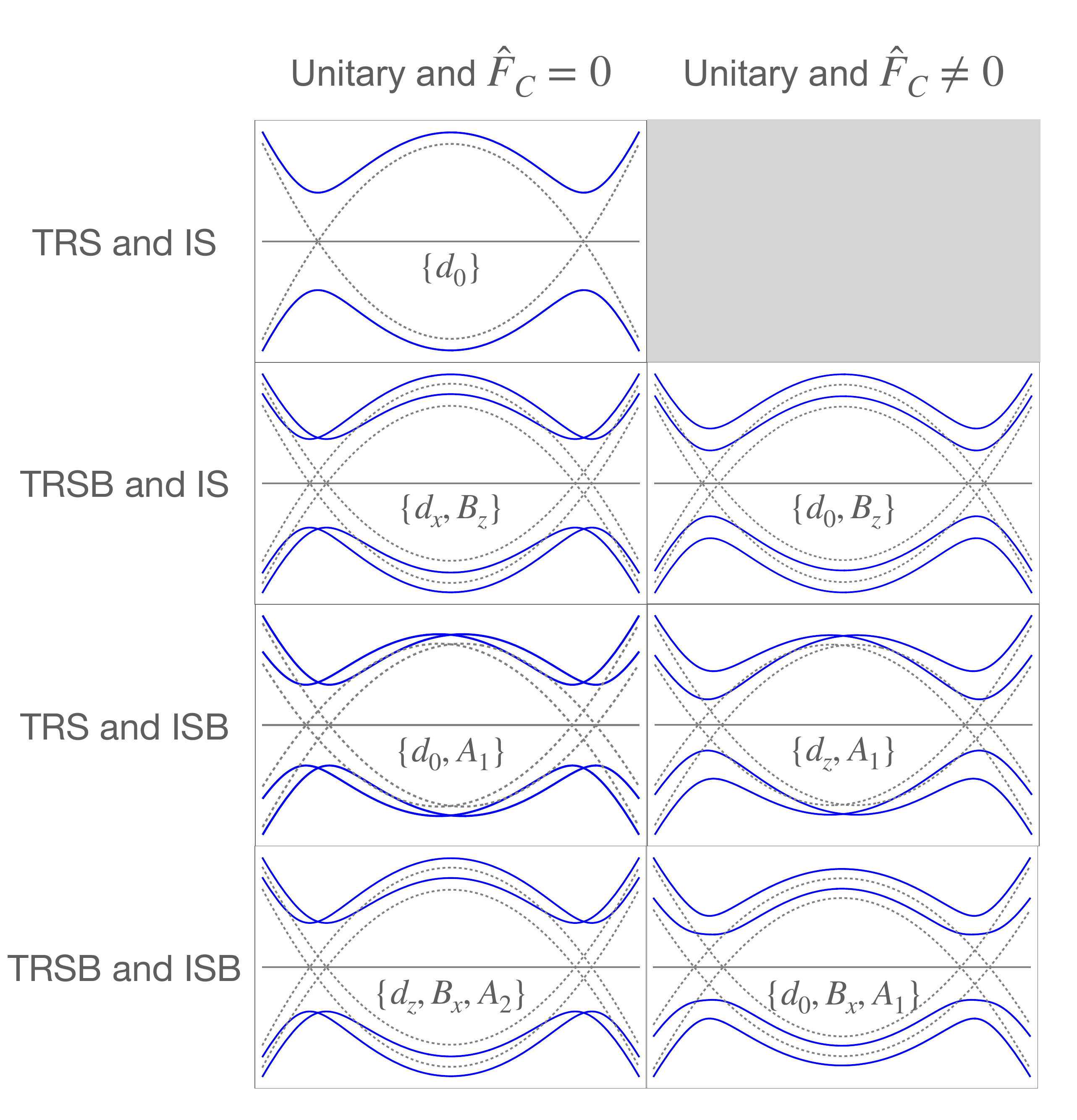}
\caption{Energy versus momentum plots of the normal and superconducting state spectra for unitary order parameters. The  momentum is taken along the $k_x$ direction. The blue lines correspond to the superconducting spectrum, while the gray dashed lines correspond to the normal state bands (both particle and hole sectors). The horizontal gray lines marks the Fermi level. Each row encodes different symmetries in the normal state Hamiltonian specified on the left. The left (right) column corresponds to cases with $\hat{F}_C(\bk)=0$ ($\hat{F}_C(\bk)\neq0$). The gray box indicates that there is no example for the respective conditions. In brackets we highlight the finite parameters: $\{d_0, d_x, d_y, d_z\}$ correspond to the d-vector parametrization of the superconducting order parameter, as given by Eq. \ref{eq:gap}, $\bb = (B_x,B_y,B_z)$ correspond to magnetic field components, and $\bg (\bk) = (-A_1 k_y, A_1 k_x , -A_2 k_z)$ correspond to Rashba ($A_1$) or Ising ($A_2$) SOC. The parameters used for the plots are the following: $\epsilon(\bk) = \bk^2 -\mu$, with $\mu=2$, $d_0=d_x=d_z=1$, and $B_x=B_z=A_1=A_2=0.25$.}
\label{fig:unitary}
\end{figure}

This result corresponds to two bands in the normal state, $\xi_\pm(\bk) = \epsilon(\bk) \pm |\mathbf{s}(\bk)|$, developing the same gap with magnitude $\Delta_U(\bk)$. From this quadrant we already recover the four eingenenergies that are expected for the four-dimensional BdG Hamiltonian [the lower right quadrant gives the same result, assuming that we do not break both TRS and IS concomitantly such that $\mathbf{s}(\bk)$ has well defined parity , or $\bg(\bk)\cdot \bb = 0$]. In the simplest case of a normal state with both TRS and IS, we find a doubly-degenerate superconducting spectrum, as depicted in the top left corner of Fig. \ref{fig:unitary}. If the normal state has ISB, TRSB, or both, the normal state spectrum is nondegenerate and the particle and hole bands cross away from the Fermi energy. This crossing is inherited by the superconducting spectrum and is protected in case $\hat{F}_C(\bk)=0$, as depicted on the three last lines of the left column of Fig. \ref{fig:unitary}.

To understand the effects of $\hat{F}_C(\bk)\neq0$, let's focus on the  scenario with a spin singlet superconductor developing out of a normal state subject to an external magnetic field along the z-direction. In this case $[\hat{H}_0(\bk)]^2$ is diagonal and $\hat{F}_C(\bk) = f_x(\bk) \hat{\sigma}_x$, omitting the momentum dependence:
\begin{eqnarray}
\hat{\mathcal{H}}_{BdG}^2 (\bk)
 &=&\begin{pmatrix} 
h_0 + h_z  & 0 &  0 & f_x\\
0 & h_0 - h_z  & f_x & 0 \\
0 & f_x^* & h_0 + h_z & 0 \\
f_x^* & 0 & 0 & h_0 - h_z &
 \end{pmatrix},
\end{eqnarray}
where $h_0(\bk) = \epsilon^2(\bk) + B_z^2$, $h_z(\bk) = 2\epsilon(\bk) B_z$, and $f_x(\bk) = 2 d_0(\bk) B_z$. Note that $f_x(\bk)$ connects different eigenvalues $E^2_\pm(\bk)$ [for the case of $\hat{F}_C(\bk)=0$]. The ultimate effect on the spectrum can be then understood by splitting the matrix above in two $2\times 2$ matrices:
\begin{eqnarray}
\hat{\mathcal{H}}_{BdG}^{2\pm} (\bk)
 &=&\begin{pmatrix} 
h_0(\bk) \pm h_z(\bk)  &  f_x(\bk) \\
 f_x^*(\bk)  & h_0(\bk)\mp h_z(\bk)  \\
 \end{pmatrix},
\end{eqnarray}
with eigenvalues $h_0(\bk) \pm \sqrt{h_z^2(\bk) + |f_x(\bk)|^2}$. This means that the original eigenvalues $E_\pm^2 = h_0(\bk) \pm h_z(\bk)$ [for $\hat{F}_C(\bk)=0$] are split. At the crossing of the particle and hole bands at high energy there is a gap opening in the superconducting state, creating a ``mirage gap" at finite energy. This aspect can also be understood by the fact that the fitness measure is written in terms of products of the normal state Hamiltonian $\hat{H}_0(\bk)$, which connects particle and particle spaces, and the superconducting gap matrix $\hat{\Delta}(\bk)$, which connects particle and hole spaces. The fitness matrices therefore connects particle and hole spaces, and can be thought of as an hybridization between these sectors, allowing for gap openings when bands associated with these different sectors cross at finite energy.

Fig. \ref{fig:unitary} displays the superconducting spectra for selected cases of unitary superconductivity, depending on the symmetries in the normal state and a zero or finite fitness matrix. The first row corresponds to the normal state with both TRS and IS. In this case the normal state bands are doubly degenerate (dashed lines for both particle and hole sectors) and $\hat{F}_C(\bk)=0$ as the normal state Hamiltonian is simply proportional to the identity matrix. A unitary order parameter opens the same gap in both bands and the superconducting spectrum is doubly degenerate (blue lines). The second row corresponds to TRSB in the normal state, introduced by an external magnetic field. The bands in the normal state are Zeeman split, and once we consider both particle and hole sectors, we note crossings at finite energy. For the example with $\hat{F}_C(\bk)=0$ we choose a triplet order parameter with a d-vector perpendicular to the magnetic field. The superconducting spectrum in this case inherits the crossing of the particle and hole bands in the normal state. For the example with $\hat{F}_C(\bk)\neq0$ we choose a singlet order parameter. Note that in this case the crossings at high energies are lifted. A similar discussion holds for the third and fourth rows, with normal state Hamiltonian breaking only IS or both IS and TRS, respectively.

\subsection{Case II: nonunitary order parameter preserving TRS}\label{sec:caseii}

The upper left block of the BdG Hamiltonian, Eq. \ref{eq:square}, now reads: 
\begin{eqnarray}\label{eq:EEFc0}
[\hat{H}_0(\bk)]^2 + \hat{\Delta}(\bk)\hat{\Delta}^\dagger(\bk) &=&  \{[\epsilon(\bk)]^2 + |\mathbf{s}(\bk)|^2 + |\Delta_U(\bk)|^2\} \hat{\sigma}_0 \\ \nonumber &&
+[ 2 \epsilon(\bk)\mathbf{s}(\bk)  + \bq_{NU}(\bk)]  \cdot \hat{\boldsymbol{\sigma}}.
\end{eqnarray}

The eigenvalues of the square of the BdG Hamiltonian for $\hat{F}_C(\bk)=0$ are then 
\begin{eqnarray}
E^2_\pm &=& [\epsilon(\bk)]^2 + |\mathbf{s}(\bk)|^2 + |\Delta_U(\bk)|^2 \pm |2 \epsilon(\bk)\mathbf{s}(\bk) + \bq_{NU}(\bk)| \\ \nonumber
&=&  [\epsilon(\bk) \pm |\mathbf{s}(\bk)|]^2+ |\Delta_U(\bk)|^2 \pm |\bq_{NU}(\bk)|  \pm  4| \epsilon(\bk)\mathbf{s}(\bk)| |\bq_{NU}(\bk)|\cos\theta,
\end{eqnarray}
where $\theta $ is the angle between $\bq_{NU}(\bk)$ and $\mathbf{s}(\bk)$. 

The eigenvalues of the BdG Hamiltonian for $\hat{F}_C(\bk)=0$ are now:
\begin{eqnarray}
\hspace{-1.5cm}E_{\pm\pm}= \pm \sqrt{ [\epsilon(\bk) \pm |\mathbf{s}(\bk)|]^2+ |\Delta_U(\bk)|^2 \pm |\bq_{NU}(\bk)|  +  4| \epsilon(\bk)\mathbf{s}(\bk)| |\bq_{NU}(\bk)|\cos\theta}.\nonumber\\
\end{eqnarray}

Under the simplifying assumption that $\bq_{NU}(\bk)$ and $\mathbf{s}(\bk)$ are perpendicular, the dispersion corresponds to two bands, $\xi_\pm(\bk) = \epsilon(\bk) \pm |\mathbf{s}(\bk)|$, developing  gaps with different magnitude $\Delta_\pm(\bk) = \sqrt{|\Delta_U(\bk)|^2 \pm |\bq_{NU}(\bk)|}$. Again, from this quadrant we already recover the four eingenenergies that are expected for the superconducting dispersion of a two band superconductor (the lower right quadrant gives the same result, assuming that we do not break both TRS and IS concomitantly).

\begin{figure}[h]
\includegraphics[width=0.8\textwidth]{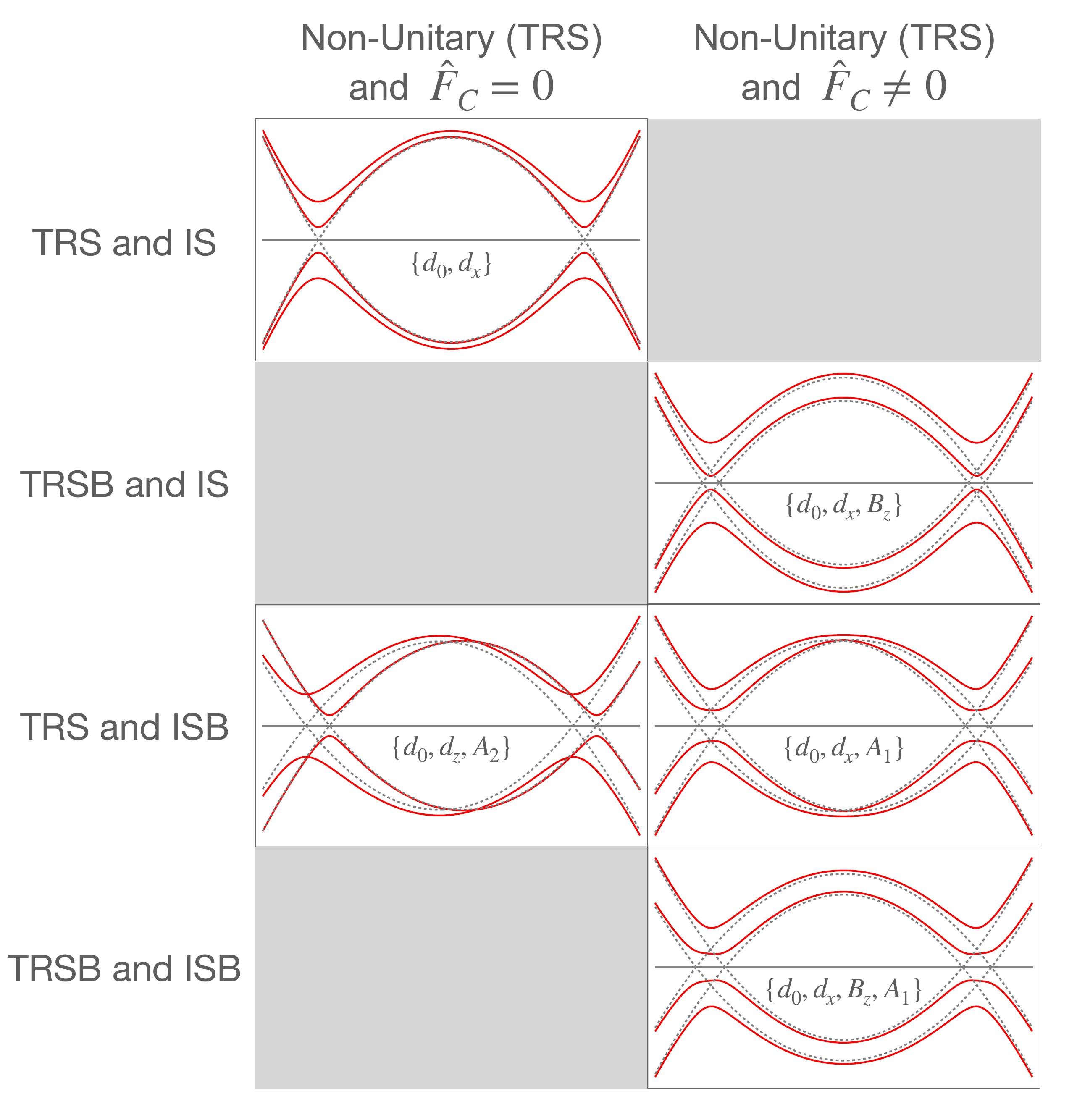}
\caption{Superconducting spectra for nonunitary order parameters preserving TRS. The red lines correspond to the superconducting spectrum, while the gray dashed lines correspond to the normal state bands (both particle and hole sectors). Same description as for Fig. \ref{fig:unitary}. The parameters used for the plots are the following: $\epsilon(\bk) = \bk^2 -\mu$, with $\mu=2$, $d_0=0.5$, $d_x=d_z=0.25$, and $B_z=A_1=A_2=0.25$. }
\label{fig:NunitaryTRS} 
\end{figure}

Fig. \ref{fig:NunitaryTRS} displays the superconducting spectra for selected cases of TRS nonunitary superconductivity. As the nonunitary order parameter is not associated with TRS, it must be associated with ISB, therefore we chose a mix of spin singlet and spin triplet states. The first row corresponds to the normal state with both TRS and IS. In this case the normal state bands are doubly degenerate (dashed lines for both particle and hole sectors) and $\hat{F}_C(\bk)=0$ as the normal state Hamiltonian is simply proportional to the identity matrix. A nonunitary order parameter opens two different gaps and the superconducting spectrum is not degenerate (red lines). Note that this is a very unlikely scenario since the order parameter spontaneously breaks inversion symmetry. The third row corresponds to ISB in the normal state, introduced by Rashba or Ising SOC. The bands in the normal state are split and there are crossings of particle and hole bands at finite energy.  For the example with $\hat{F}_C(\bk)= 0$ we choose the SOC vector to be parallel to the d-vector and we find that the superconducting spectrum displays band crossings. For the case with $\hat{F}_C(\bk) \neq 0$ we choose a SOC with a component in the same direction as the d-vector. Note that in this case the crossings of particle and hole bands at finite energies are lifted. A similar discussion holds for the second and fourth rows, with normal state Hamiltonian breaking only TRS or both IS and TRS, respectively. Note that as we have a mix of singlet and triplet states, once TRS is broken we necessarily have a finite superconducting fitness.

\subsection{Case III: Nonunitary order parameter breaking TRS}\label{sec:caseiii}

The discussion concerning the superconducting spectrum for $\hat{F}_C(\bk)=0$ is similar to the one provided for Case II above. 

\begin{figure}[h]
\includegraphics[width=0.8\textwidth]{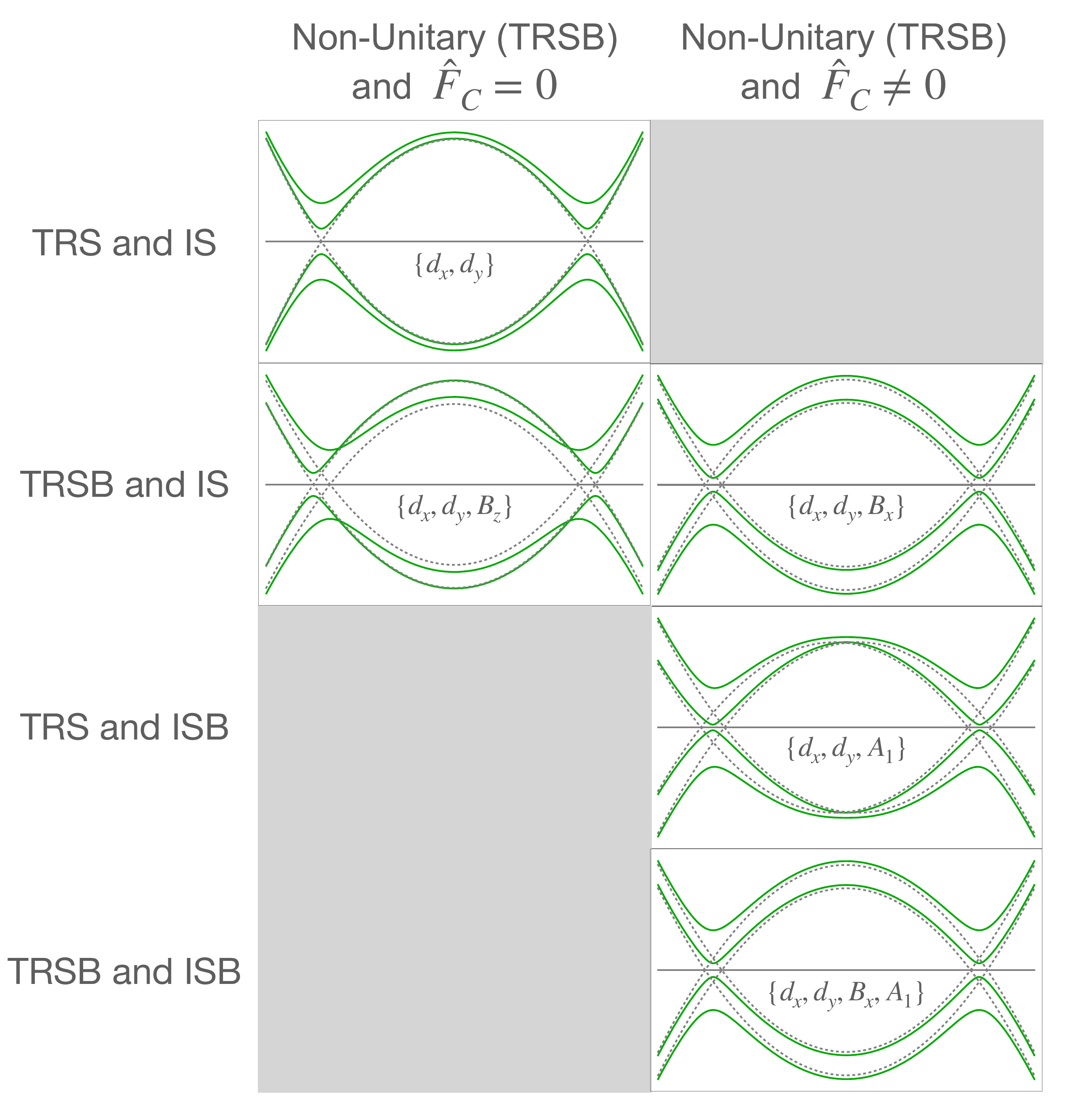}
\caption{Superconducting spectra for nonunitary order parameters breaking TRS. The green lines correspond to the superconducting spectrum, while the gray dashed lines correspond to the normal state bands (both particle and hole sectors). Same description as for Fig. \ref{fig:unitary}. The parameters used for the plots are the following: $\epsilon(\bk) = \bk^2 -\mu$, with $\mu=2$, $d_x=0.5$, $d_y=i 0.25$, and $B_z=A_1=A_2=0.25$.  }
\label{fig:NunitaryTRSB} 
\end{figure}

Fig. \ref{fig:NunitaryTRSB} displays the superconducting spectra for selected cases of TRSB nonunitary superconductivity. The first row corresponds to the normal state with both TRS and IS. As in the last case, a nonunitary order parameter opens two different gaps and the superconducting spectrum is not degenerate (green lines).  The second row corresponds to TRSB in the normal state, introduced by an external magnetic field. The bands in the normal state are Zeeman split, and once we consider both particle and hole sectors, we note crossings at finite energy. For the example with $\hat{F}_C(\bk)=0$ we choose a triplet order parameter with a complex multi-component d-vector perpendicular to the magnetic field. The superconducting spectrum in this case inherits the crossing of the particle and hole bands in the normal state. For the example with $\hat{F}_C(\bk)\neq0$ we choose a complex multi-component d-vector with a component parallel to the magnetic field. Note that in this case the crossings at high energies are lifted. A similar discussion holds for the third and fourth rows, with normal state Hamiltonian breaking only IS or both IS and TRS, respectively.

From the discussion above, we conclude that for simple superconductors with only the spin as an internal DOF: i) crossings in the superconducting spectra are inherited from the normal state spectra, considering both particle and hole sectors; ii) the crossings are protected in the presence of symmetry breaking fields with  $\hat{F}_C(\bk)=0$; iii) the crossings are lifted if $\hat{F}_C(\bk)\neq 0$. Therefore the opening of gaps at high energies is a feature primarily associated with a finite fitness measure, $\hat{F}_C(\bk)\neq 0$, and not  with the nonunitary aspect of the superconducting state.

\section{Superconductivity in complex quantum materials}\label{sec:complex}

The discussion of nonunitary order parameters becomes much richer if we move to scenarios with more than one internal degree of freedom. The first non-trivial scenario appears considering models with two internal DOFs, which can be associated with two orbitals, or a sublattice structure. The BdG Hamiltonian for the two orbital scenario has the same form as the one displayed in Eq. \ref{Eq:BdG}, with $n=2$.

The normal state Hamiltonian can be generally written as:
\begin{eqnarray}
\hat{H}_0(\bk)  = \sum_{ab} h_{ab}(\bk) \hat{\tau}_a \otimes \hat{\sigma}_b,
\end{eqnarray}
where $h_{ab}(\bk)$ are real functions of momenta encoding all the information about hopping amplitudes and spin-orbit coupling (SOC). $\hat{\tau}_a$ and $\hat{\sigma}_a$ are Pauli matrices for $a=\{1,2,3\}$ or the $2\times 2$ identity matrix for $a=0$, encoding the orbital or sublattice and spin DOFs, respectively. Time-reversal symmetry is now implemented by $\hat{\Theta} = \mathcal{K} \hat{\tau}_0 \otimes (i\hat{\sigma}_2)$, accompanied by the change in momenta $\bk\rightarrow -\bk$, leaving the orbital and sublattice DOFs invariant. Inversion symmetry can appear in different flavours, depending on the nature of the new DOF. In case the new DOF is associated with two orbitals with same parity, inversion is implemented as $\hat{P} = \pm \hat{\tau}_0 \otimes \hat{\sigma}_0$, in case the orbitals have opposite parity, inversion is implemented as $\hat{P} = \pm \hat{\tau}_3 \otimes \hat{\sigma}_0$, and in case they are associated with a sublattice structure, as $\hat{P} = \pm \hat{\tau}_1 \otimes \hat{\sigma}_0$, such that inversion exchanges sublattices. In presence of both inversion and time-reversal symmetries, the set of symmetry allowed pairs $(a,b)$ is reduced from a total of 16 to only 5 plus the identity matrix, labelled as $(0,0)$. The allowed pairs for each case are summarized in Table \ref{Tab:H0}. In all cases the set of matrices $(a,b)$ [excluding $(0,0)$] forms a totally anti-commuting set of matrices, similar in structure to the Pauli matrices.

\begin{table}[h]
\begin{center}
    \begin{tabular}{| c | c | c | c | c | }
    \hline
    $(a,b)$ &   $\bk$ & EP & OP & SL    \\ \hline
    $(0,0)$ &  Even  & $\checkmark$  & $\checkmark$ & $\checkmark$ \\ \hline
    $(0,1)$ &  Odd   & & & \\ \hline
    $(0,2)$ &  Odd   & & & \\ \hline
    $(0,3)$ &  Odd   & & & \\ \hline
    $(1,0)$ &  Even  & $\checkmark$ & & $\checkmark$  \\ \hline
    $(1,1)$ &  Odd   & & $\checkmark$ & \\ \hline
    $(1,2)$ &  Odd   & & $\checkmark$ & \\ \hline
    $(1,3)$ &  Odd   & & $\checkmark$ & \\ \hline
    $(2,0)$ &  Odd   & & $\checkmark$ & $\checkmark$\\ \hline
    $(2,1)$ &  Even  & $\checkmark$ & & \\ \hline
    $(2,2)$ &  Even  & $\checkmark$ & & \\ \hline
    $(2,3)$ &  Even  & $\checkmark$ & & \\ \hline
    $(3,0)$ &  Even  & $\checkmark$ & $\checkmark$ & \\ \hline
    $(3,1)$ &  Odd   & & & $\checkmark$ \\ \hline
    $(3,2)$ &  Odd   & & & $\checkmark$ \\ \hline
    $(3,3)$ &  Odd   & & & $\checkmark$ \\ \hline
    \end{tabular}
        \end{center}
    \caption{  \label{Tab:Parity} Symmetry allowed $(a,b)$ terms in the normal state Hamiltonian under time-reversal symmetry and inversion symmetry for different scenarios: equal parity orbitals (EP), opposite parity orbitals (OP), and sublattice structure (SL). The second column indicates if the respective $h_{ab}(\bk)$ is an even or an odd function of momenta.}
    \label{Tab:H0}
\end{table}




The most general order parameter can now be written as:
\begin{eqnarray}
\hat{\Delta}(\bk) = \sum_{[a,b]} d_{ab}(\bk) \hat{\tau}_a \otimes \hat{\sigma}_b (i \hat{\sigma}_2).
\end{eqnarray}
Here the sum runs over the pairs $[a,b]$ in the same irreducible representation (a more refined discussion based on point group symmetries for d-electrons is given in Sec. \ref{sec:delectrons}). Time-reversal operation applied to the order parameter gives us $\hat{\Theta} \hat{\Delta}(\bk)\hat{\Theta}^{-1}  =  \sum_{[a,b]} d_{ab}^*(\bk) \hat{\tau}_a \otimes \hat{\sigma}_b (i \hat{\sigma}_2)$, such that if all components $d_{ab}$ have the same phase, the order parameter is time-reversal invariant [up to a $U(1)$ transformation]. 

Note that given fermionic anti-symmetry, $\hat{\Delta}(\bk) = -\hat{\Delta}^\text{T}(-\bk)$, all $d_{ab}(\bk)$ accompanied by (anti-)symmetric matrices $\hat{\tau}_a \otimes \hat{\sigma}_b (i \hat{\sigma}_2)$ must be (even) odd in $\bk$. This feature allows us to summarize the basic properties concerning momentum dependence and parity of all possible order parameter components in two-DOF models in Table \ref{Tab:SC}.

We can now evaluate the gauge-invariant composition of the order parameter (omitting the momentum dependence for a more concise notation):
\begin{eqnarray}
\hat{\Delta} \hat{\Delta}^\dagger = \Delta_U^2 \hat{\tau}_0\otimes\hat{\sigma}_0 + q_{NU}^{ab} \hat{\tau}_a\otimes\hat{\sigma}_b,
\end{eqnarray}
where we define, in analogy to the case of simple superconductors discussed above:
\begin{eqnarray}
\Delta_U^2=\sum_{[a,b]}|d_{ab}|^2
\end{eqnarray}
as the magnitude of the unitary part of the gap, and 
 \begin{eqnarray}
 q_{NU}^{ab}&=& Tr[\hat{\Delta} \hat{\Delta}^\dagger \hat{\tau}_a\otimes\hat{\sigma}_b]/4 \\ \nonumber
 &=& \sum_{[a,b] \neq [0,0]} [d_{00}d_{ab}^*+d_{00}^*d_{ab}]\\ \nonumber
 &&+\sum_{\{[m,n], [p,q]\}\neq [0,0]}  [d_{mn}d_{pq}^*] Tr[\hat{\tau}_m \hat{\tau}_p \hat{\tau}_a] Tr[\hat{\sigma}_n \hat{\sigma}_q\hat{\sigma}_b]/4,
 \end{eqnarray}
the nonunitary component of the order parameter. Note that now non-unitarity can be associated with both spin and orbital polarization, with the finite expectation value $\langle \hat{\Delta}^\dagger \hat{\tau}_a \otimes \hat{\sigma}_b \hat{\Delta}\rangle$ for a given momenta, which can have a finite average over the Fermi surface. 

In analogy to the single band scenario, we can again define a time-reversal symmetry odd gauge-invariant product \cite{Brydon:2018}:
\begin{eqnarray}
\hat{\Delta}(\bk)\hat{\Delta}^\dagger(\bk)  - \hat{\Delta}_T(\bk)\hat{\Delta}_T^\dagger(\bk) = \sum_{[a,b]} q^{ab}_{TRO}(\bk) \hat{\tau}_a \otimes \hat{\sigma}_b.
\end{eqnarray} 
If there is at least one $q^{ab}_{TRO}(\bk) \neq 0$, the TRSB order parameter develops a spin-orbital polarization.

Given the greater number of order parameter basis matrixes, there are many more possibilities to generate nonunitary superconductivity. In contrast to the single band scenario, these are not necessarily associated with symmetry breaking.

\begin{table}[h]
\begin{center}
    \begin{tabular}{| c | c | c | c | c | c |}
    \hline
    $[a,b]$ &   $\hat{\tau}_a\otimes \hat{\sigma}_b (i\sigma_2)$ & $d_{ab}(\bk)$ & EP & OP & SL   \\ \hline
    $[0,0]$ &   A & E & E & E & E \\ \hline
    $[0,1]$ &   S & O & O & O & O \\ \hline
    $[0,2]$ &   S & O & O& O & O \\ \hline
    $[0,3]$ &   S & O & O & O & O \\ \hline
    $[1,0]$ &   A & E & E & O & E \\ \hline
    $[1,1]$ &   S & O & O & E & O \\ \hline
    $[1,2]$ &   S & O & O & E & O \\ \hline
    $[1,3]$ &   S & O & O & E & O \\ \hline
    $[2,0]$ &   S & O & O & E & E \\ \hline
    $[2,1]$ &   A & E & E & O & O \\ \hline
    $[2,2]$ &   A & E & E & O & O \\ \hline
    $[2,3]$ &   A & E & E & O & O \\ \hline
    $[3,0]$ &   A & E & E & E & O \\ \hline
    $[3,1]$ &   S & O & O & O & E \\ \hline
    $[3,2]$ &   S & O & O& O & E \\ \hline
    $[3,3]$ &   S & O & O & O & E \\ \hline
    \end{tabular}
        \end{center}
    \caption{  \label{Tab:SC} Properties of the superconducting order parameters. The first column gives the label $[a,b]$, the second column indicates if the matrix  $\hat{\tau}_a\otimes \hat{\sigma}_b (i\sigma_2)$ is symmetric (S) or anti-symmetric (A). The third column indicates if the accompanying function $d_{ab}(\bk)$ is even (E) or odd (O) in momentum $\bk$. The last three columns indicate the parity even (E) or odd (O) for different scenarios: equal parity orbital (EP), opposite parity orbitals (OP), and sublattice structure (SL). The parity indicated in the three last columns indicate the global parity of the order parameter, combining the parity of the matrix structure and of the function $d_{ab}(\bk)$.}
    \label{tab:parity}
\end{table}

\subsection{Inversion symmetry}\label{sec:inversion}

If only inversion symmetry is present, the order parameters are split in two sectors of distinct parity. We can gather the following possible superpositions for each scenario, omitting the momentum dependence of the $d_{ab}(\bk)$ functions:

\begin{eqnarray}
\hat{\Delta}_{EP}^{Even} &=& [d_{00} \hat{\tau}_0\otimes\hat{\sigma}_0
+ d_{10} \hat{\tau}_1\otimes\hat{\sigma}_0+d_{30} \hat{\tau}_3\otimes\hat{\sigma}_0\\ \nonumber&&+d_{21} \hat{\tau}_2\otimes\hat{\sigma}_1+d_{22} \hat{\tau}_2\otimes\hat{\sigma}_2+d_{23} \hat{\tau}_2\otimes\hat{\sigma}_3]i\hat{\sigma}_2
\end{eqnarray}

\begin{eqnarray}
\hat{\Delta}_{OP}^{Even} &=& [d_{00} \hat{\tau}_0\otimes\hat{\sigma}_0
+d_{30} \hat{\tau}_3\otimes\hat{\sigma}_0]i\hat{\sigma}_2\\
\hat{\Delta}_{OP}^{Odd} &=& [ d_{10} \hat{\tau}_1\otimes\hat{\sigma}_0+d_{21} \hat{\tau}_2\otimes\hat{\sigma}_1+d_{22} \hat{\tau}_2\otimes\hat{\sigma}_2+d_{23} \hat{\tau}_2\otimes\hat{\sigma}_3]i\hat{\sigma}_2
\end{eqnarray}

\begin{eqnarray}
\hat{\Delta}_{SL}^{Even} &=& [d_{00} \hat{\tau}_0\otimes\hat{\sigma}_0
+ d_{10} \hat{\tau}_1\otimes\hat{\sigma}_0]i\hat{\sigma}_2\\
\hat{\Delta}_{SL}^{Odd} &=& [d_{30} \hat{\tau}_3\otimes\hat{\sigma}_0+d_{21} \hat{\tau}_2\otimes\hat{\sigma}_1+d_{22} \hat{\tau}_2\otimes\hat{\sigma}_2+d_{23} \hat{\tau}_2\otimes\hat{\sigma}_3]i\hat{\sigma}_2
\end{eqnarray}

Note that, even in presence of inversion symmetry, all scenarios above can potentially host nonunitary order parameters. Below we refine the discussion considering point group symmetries, and show that in certain cases multiple realizations of order parameters can be found in the trivial irrep, such that no extra symmetry is broken.

\section{Application to d-electron systems}\label{sec:delectrons}

Here we focus on d-electron systems, belonging to the EP scenario discussed above. The conclusions drawn here can be applied to multiple families of materials (ruthenates, pnictides, and transition metal dichalcogenides, to name a few), and can be easily generalized to scenarios of orbitals with OP or with a SL structure. 

In order to give a concrete example, in Appendix A we explicitly derive the symmetry classification of the order parameters for the $D_{2h}$ point group for different choices of pairs of d-electrons. The derivations for other point groups follow the very same lines. The results for point groups  $D_{2h}$, $D_{4h}$ and $D_{6h}$ with inversion symmetry and $C_{2v}$, $D_{2d}$ and $C_{6v}$ without inversion symmetry are shown in Tabs. \ref{tab:d2h} - \ref{tab:c6v} in Appendix B. Interestingly, these groups are associated with several materials that develop TRSB SC, according to Tab. I in Ref. \cite{Ghosh:2020}. In absence of inversion symmetry orbitals of different parity can mix, but here we assume that the mixing is small and explore the discussion only with d-electrons even in the case if ISB in the normal state. 

For the classification of the order parameters labelled as $[a,b]$ according to the irreducible representations of the point group, we write the point group operations as unitary transformations acting as $\hat{\Delta}(\bk) \rightarrow \hat{\Delta}'({\bk'}) = U \hat{\Delta}({\bk'}) U^T$, where the superscript $T$ indicates the transpose and $\bk' = U\bk$ is the rotated $\bk$ vector. Note that since the inversion operation acts trivially on the matrix structure of the order parameters, their parity in the EP scenario is directly determined by the parity of the accompanying momentum dependent function $d_{ab}(\bk)$, which is pre-determined by fermionic antisymmetry (see Table \ref{tab:parity}). 

From here on, we focus on order parameters that are momentum independent, restricting the analysis to even parity order parameters. In this case, the symmetry properties of the order parameters are completely determined by their matrix structure. Table \ref{tab:irreps} summarizes how each matrix $\hat{\tau}_a \otimes \hat{\sigma}_b (i \hat{\sigma}_2)$ transforms under the point group operations (see Appendix A), allowing for the identification of the irreducible representations in accordance with the character table  (Tab. \ref{tab:character}), as summarized in the right column of Tab. \ref{tab:irreps}.

\begin{table}[h]
\begin{center}
    \begin{tabular}{| c | c | c | c | c | c |}
    \hline
$[a,b]$ & $E$ & $C_{2z}$ & $C_{2x}$ & $C_{2y}$ & irrep \\ \hline
$[0,0]$ & 1 & 1 & 1 & 1 & $\mathbf{A_{1g}}$ \\ \hline
$[1,0]$ & 1 & 1 & -1 & -1 & $B_{1g}$ \\ \hline
$[3,0]$ & 1 & 1 & 1 & 1 & $\mathbf{A_{1g}}$ \\ \hline
$[2,1]$ & 1 & -1 & -1 & 1 & $B_{2g}$ \\ \hline
$[2,2]$ & 1 & -1 & 1 & -1 & $B_{3g}$ \\ \hline
$[2,3]$ & 1 & 1 & 1 & 1 & $\mathbf{A_{1g}}$ \\ \hline
    \end{tabular}
        \end{center}
    \caption{ Identification of the irreps of the order parameters for $\{d_{xz}, d_{yz}\}$ orbitals in the $D_{2h}$ point group. Highlighted in bold are the three realizations of order parameters with $A_{1g}$ symmetry.}
    \label{tab:irreps} 
\end{table} 

In the Appendix B we provide tables compiling the results for multiple point group symmetries and different choices of pairs of d-orbitals (Tabs. \ref{tab:d2h} - \ref{tab:c6v}). The bottom line is that there are always two or more basis matrices that transform according to the trivial representation ($A_{1g}$ or $A_1$, for groups with and without inversion symmetry, respectively): $\{[0,0],[3,0]\}$, $\{[0,0],[2,3]\}$, or $\{[0,0],[3,0],[2,b]\}$, for $b=1,2,3$. For these scenarios, the gap can be a linear superposition of the basis matrices and follows (note that there is no sum over the index $b$):
\begin{eqnarray}
\hat{\Delta} (\bk) &=& [d_{00}(\bk) \hat{\tau}_0 \otimes \hat{\sigma}_0 + d_{30}(\bk)\hat{\tau}_3 \otimes \hat{\sigma}_0  + d_{2b}(\bk)\hat{\tau}_2\otimes\hat{\sigma}_b](i\hat{\sigma}_2)
\end{eqnarray}
such that, omitting the $\bk$ dependence:
\begin{eqnarray}
\hat{\Delta} \hat{\Delta}^\dagger &=& (|d_{00}|^2+ |d_{30}|^2 + |d_{2b}|^2)  \hat{\tau}_0 \otimes \hat{\sigma}_0 
\\ \nonumber
&+& (d_{00}d_{30}^* + d_{00}^*d_{30}) \hat{\tau}_3 \otimes \hat{\sigma}_0 + (d_{00}d_{2b}^* +d_{00}^*d_{2b}) \hat{\tau}_2 \otimes \hat{\sigma}_b
\\ \nonumber
&+&i(d_{2b}d_{30}^*-d_{2d}^*d_{30})\hat{\tau}_1\otimes\hat{\sigma}_b,
\end{eqnarray}

Note that, in case all $d_{ab}$ are real (the order parameter is TRS), the matrices which appear in $\hat{\Delta} \hat{\Delta}^\dagger$ are necessarily in the trivial representation ($\hat{\tau}_0 \otimes \hat{\sigma}_0 , \hat{\tau}_3 \otimes \hat{\sigma}_0,  \hat{\tau}_2 \otimes \hat{\sigma}_b$). This reflects the fact that non-unitarity in multi-orbital superconductors (as defined by a finite $q_{NU}^{ab}(\bk)$) is not necessarily associated with symmetry breaking, in contrast to the discussion in the single orbital scenario. If the order parameter breaks TRS and develops a finite $q_{TRO}^{ab}(\bk)$, there is a finite spin-orbital polarization of the pairs. In the next section we are going to discuss how these aspects are reflected in the superconducting energy spectra.

\section{Signatures of nonunitary order parameters in the two orbital scenario}\label{sec:example}

Now we translate some of the conclusions we found for the case of simple superconductors, carrying only the spin as an internal DOF, to the more complex scenario of superconductors with an extra orbital DOF.

In the normal state, in presence of inversion and time-reversal symmetries, the spectrum is constituted of two doubly-degenerate bands. The normal state spectrum in presence of TRS and IS can be obtained by diagonalizing $\hat{H}_0(\bk)$, leading to:
\begin{eqnarray}
\xi_\pm (\bk) = h_{00}(\bk) \pm |\bh (\bk)|,
\end{eqnarray}
where $\bh (\bk)$ is the vector formed by $h_{ab}(\bk)$ for $(a,b)\neq (0,0)$, in analogy to the $\bs(\bk)$ vector for the single DOF case discussed above. For concreteness, here we take the example of d-electrons $\{d_{xz}, d_{yz}\}$ and point group $D_{2h}$. We write the parameters $h_{ab}(\bk)$ as expansions around the $\Gamma$ point:
\begin{eqnarray}
h_{00}(\bk) &=& |\bk|^2/(2m_1) - \mu_1,\\ \nonumber
h_{10}(\bk) &=& \beta k_x k_y,\\ \nonumber
h_{21}(\bk) &=& \gamma_1 k_x k_z,\\ \nonumber
h_{22}(\bk) &=& \gamma_2 k_yk_z,\\ \nonumber
h_{23}(\bk) &=& \alpha,\\ \nonumber
h_{30}(\bk) &=& |\bk|^2/(2m_2) - \mu_2,
\end{eqnarray}
here $h_{00}(\bk)$ and $h_{30}(\bk)$ correspond to intra-orbital hopping, $h_{10}(\bk)$ to inter-orbital hopping, $h_{23}(\bk)$ to atomic SOC, and $h_{21,22}(\bk)$ to $\bk$-dependent SOC. For the plots below we choose the parameters $(m_1,m_2,\mu_1,\mu_2,\alpha,\beta,\gamma_1,\gamma_2) = (0.5, -2.5, 1.2, -0.5, 0.1,0.15,0.12,.0.17)$. For the discussion in absence of SOC we take $\alpha = \gamma_1=\gamma_2=0$. The parameters were chosen such that in absence of SOC the two bands in the normal state cross away from the Fermi energy.

For the discussion of the spectrum in the superconducting state, it is again useful to consider the square of the BdG Hamiltonian, which has the same structural form as Eq. \ref{eq:square}. In case $\hat{F}_C(\bk)=0$, the square of the BdG Hamiltonian is block diagonal. The top left block reads:
\begin{eqnarray}\label{eq:dispersiontwo}
[\hat{H}_0(\bk)]^2 + \hat{\Delta}(\bk)\hat{\Delta}^\dagger(\bk) &=&  \{[h_{00}(\bk)]^2 + [\bh(\bk)]^2 + |\Delta_U(\bk)|^2\} \hat{\sigma}_0 \\ \nonumber &&
+\sum_{a, b}{\vphantom{\sum}}' \left[2 h_{00}(\bk)h_{ab}(\bk) + q_{NU}^{ab}(\bk)\right] \hat{\tau}_a\otimes \hat{\sigma}_b,
\end{eqnarray}
where the primed sum in the last term excludes the case with both $a=0$ and $b=0$. Note that the square of the dispersion for the two-orbital model, given by Eq. \ref{eq:dispersiontwo}, and for the single orbital scenario with external symmetry breaking fields, given by Eq. \ref{eq:EEFc0}, have the same structure. This indicates that we can build up the discussion for the two-orbital scenario on similar lines.

Starting with unitary order parameters, the eigenenergies of the square of the BdG Hamiltonian simplify to:
\begin{eqnarray}
E_\pm^2 &=&  [h_{00}(\bk) \pm \bh(\bk)]^2 + |\Delta_U(\bk)|^2,
\end{eqnarray}
such that the eigenenergies of the BdG Hamiltonian are
\begin{eqnarray}
E_{\pm\pm}^2 &=&\pm \sqrt{  [h_{00}(\bk) \pm |\bh(\bk)|]^2 + |\Delta_U(\bk)|^2},
\end{eqnarray}
which are doubly degenerate. As in the case of a single DOF, we find two bands, $\xi_\pm (\bk) = h_{00}(\bk) \pm |\bh(\bk)|$, that develop gaps of same magnitude. 

For nonunitary order parameters, the eigenenergies of the square of the BdG Hamiltonian are:
\begin{eqnarray}
E_\pm^2 &=&  [h_{00}(\bk)]^2  + [\bh(\bk)]^2 + |\Delta_U(\bk)|^2 \pm \sum_{a,b} {\vphantom{\sum}}'  |2 h_{00}(\bk)h_{ab}(\bk) + q_{NU}^{ab}(\bk)|.
\end{eqnarray}

In case the vectors formed by the components $h_{ab}(\bk)$ and $q_{NU}^{ab}(\bk)$ are orthogonal, the square of the dispersion simplifies to:
\begin{eqnarray}
E_\pm^2 &=&  [h_{00}(\bk)  \pm \bh(\bk)]^2 + |\Delta_U(\bk)|^2 \pm \sum_{a,b} {\vphantom{\sum}}'  | q_{NU}^{ab}(\bk)|,
\end{eqnarray}
and the eigenenergies of the BdG Hamiltonian are
\begin{eqnarray}
E_{\pm\pm}^2 &=&\pm \sqrt{  [h_{00}(\bk) \pm |\bh(\bk)|]^2 + |\Delta_U(\bk)|^2\pm \sum_{a,b} {\vphantom{\sum}}'  | q_{NU}^{ab}(\bk)|}.
\end{eqnarray}
Now the two doubly-degenerate bands, $\xi_\pm (\bk) = h_{00}(\bk) \pm |\bh(\bk)|$, develop distinct gaps $|\Delta_\pm(\bk)|^2 = |\Delta_U(\bk)|^2\pm \sum_{a,b} {\vphantom{\sum}}'  | q_{NU}^{ab}(\bk)|$.

\begin{figure}[h]
\includegraphics[width=0.8\textwidth]{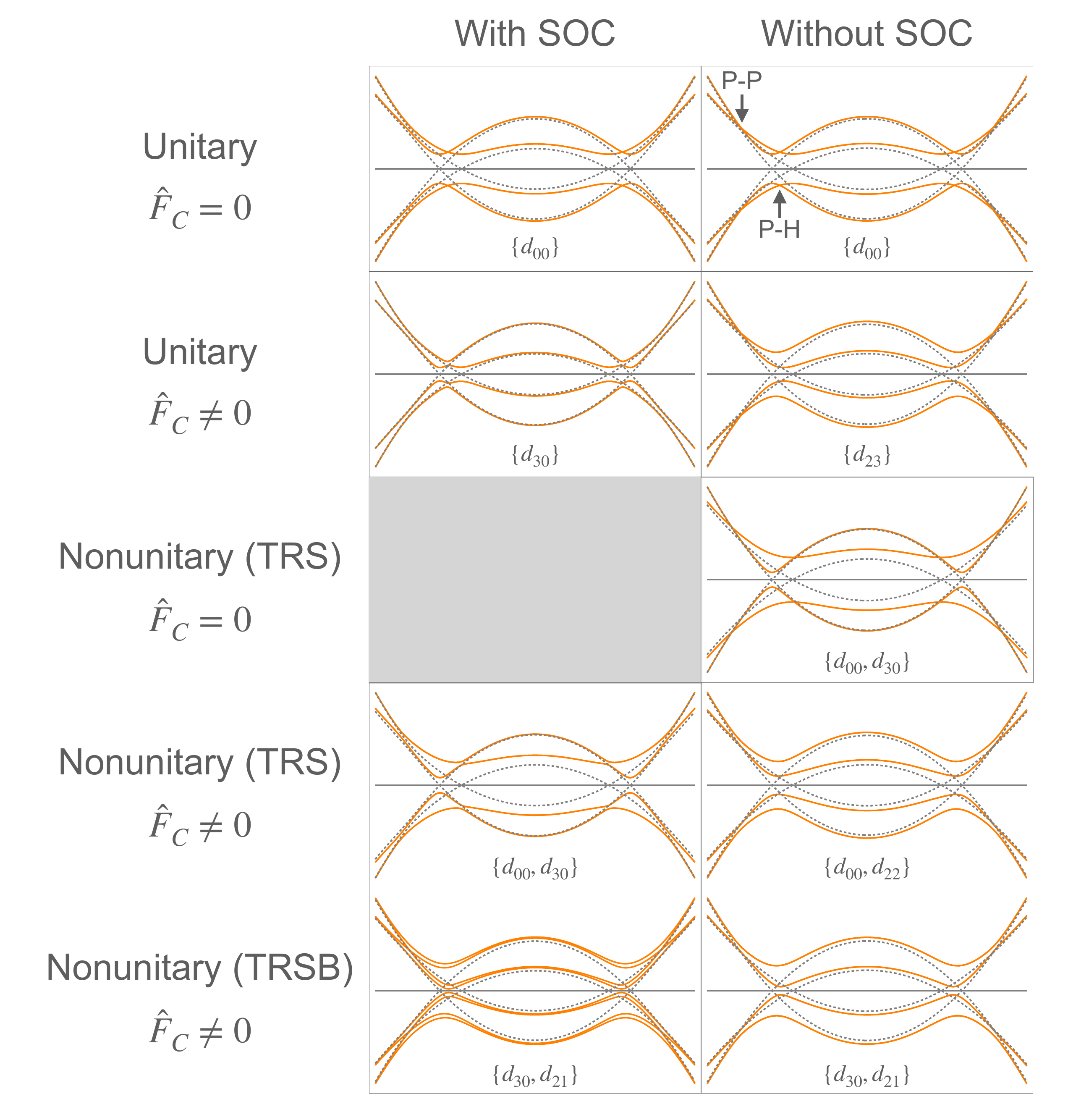}
\caption{Superconducting spectra for order parameters in the two-orbital models with TSR and IS along the $k_x$ axis. The orange lines correspond to the superconducting spectrum, while the gray dashed lines correspond to the normal state bands (both particle and hole sectors). The horizontal gray line corresponds to the `fermi level.  Note that the particle and hole bands cross each other at high energies (P-H indicated by an arrow in top right panel), as we discussed in the single DOF scenario. In addition, the bands cross each other within the particle (hole) sectors (P-P indicated by an arrow in the top right panel),  giving rise to a second type of crossing at high energies. It is indicated in each row if the order parameter is unitary or nonunitary, and if the fitness matrix is zero of non-zero. The left columns corresponds to the scenario with SOC, and the right column corresponds to the fine-tuned scenario without SOC. The gray box indicates that there is no example for these specific conditions. In brackets we highlight the finite parameters among $\{d_{00}, d_{10}, d_{21}, d_{22}, d_{23}, d_{30}\}$. When fitnite, these parameters acquire the following value: $d_{00}=0.5$, $d_{30}=0.25$, $d_{21}=d_{23}= i 0.25$, and $d_{22}=0.25$.}
\label{fig:multi}
\end{figure}

Fig. \ref{fig:multi} illustrates selected scenarios for the superconducting spectrum for unitary and nonunitary order parameters, with zero or finite fitness matrix. On the left column we have the cases with SOC, in which case the normal state bands do not cross, but there are crossings associated with the superposition of particle and hole bands (P-H crossings). On the right column we have the scenario without SOC, in which case there are crossings within the normal state bands (P-P crossings), in addition to the crossings associated with the superposition of particle and hole bands. In the first row we have examples of spectra corresponding to unitary order parameters and $\hat{F}_c(\bk) = 0$. Note that both types of crossings are preserved in the superconducting state. In the second row, once we take $\hat{F}_c(\bk) \neq 0$, the P-H crossings are lifted.  In the third row we have an example of a TRS nonunitary state ($\bq_{TRO}(\bk)= 0$) with $\hat{F}_c(\bk) = 0$. Note that this configuration is rather special as depends on setting all SOC terms to zero and is valid only for certain directions in momentum space. What is important to note here is that both types of crossings are preserved in case $\hat{F}_c(\bk) = 0$ even for a nonunitary order parameter. In the fouth row we have examples of spectra for TRS nonunitary states ($\bq_{TRO}(\bk)= 0$) with finite fitness matrix. For the scenario without SOC we see that the P-P crossings are lifted. In the fifth row we display spectra for TRSB nonunitary superconducting states ($\bq_{TRO}(\bk)\neq 0$). The spectrum in presence of SOC is more complex due to the degeneracy breaking associated with the TRSB nonunitary order parameter. For the fine-tuned scenario without SOC we do not see the splitting due to the choice of momentum along the $k_x$ direction. Any other direction away from the symmetry axes would display the splitting, as expected. For both cases, with and without SOC, the spectrum does not display crossings of the P-H or P-P types at finite energy, in agreement with the findings above for a scenario with $\hat{F}_c(\bk) \neq 0$. 

These results confirm what we had already discussed within the single DOF scenario: a finite fitness matrix is a necessary condition for the opening of gaps associated with P-H crossings. For the P-P crossings intrinsic to the band structure, the minimal requirement for their lifting is a nonunitary order parameter \emph{and} a finite fitness matrix. In conclusion, the opening of gaps at energies away from the Fermi level is primarily associated with a finite superconducting fitness. In case the crossing was already present in the normal state electronic structure, it indicates a nonunitary order parameter. These results are summarized in Table \ref{tab:conclusion}.

\begin{table}[h]
\begin{center}
    \begin{tabular}{| c | c | c | c |}
    \hline
    \multicolumn{2}{|c|}{Conditions} & P-H & P-P \\ \hline
Unitary & $\hat{F}_c(\bk) = 0$ & cross & cross \\ \hline
Unitary & $\hat{F}_c(\bk) \neq 0$ & open & cross \\ \hline
Nonunitary & $\hat{F}_c(\bk) = 0$ & cross & cross \\ \hline
Nonunitary & $\hat{F}_c(\bk) \neq 0$ & open & open \\ \hline
    \end{tabular}
    \caption{Summary of the band crossings in the superconducting state for different conditions. P-H corresponds to the superposition of particle and hole bands and P-P to the superposition of particle bands in the two DOF scenario.}
    \label{tab:conclusion}
\end{center}
\end{table}

\section{Conclusion}\label{sec:conclusion}

In this work we have revisited the notion of nonunitary order parameters. We have started with simple superconductors, emerging from electronic states with only the spin as an internal DOF. We highlight the fact that nonunitary superconducting states are necessarily associated with either ISB or TRSB of the superconducting order parameter and with the development of a two-gap structure. Furthermore, we discuss how external symmetry breaking fields in the normal state change the spectra in the superconducting state. In particular, we find that symmetry breaking terms in the normal state lead to crossings of the particle and hole bands at energies away from the Fermi energy, and that these crossings are inherited by the superconducting state in case the fitness matrix is zero, $\hat{F}_c(\bk) = 0$. Once the fitness matrix is nonzero, $\hat{F}_c(\bk) \neq 0$, these crossings are lifted.

With this more refined understanding of the superconducting spectra in simple superconductors, we moved to the minimal scenario to treat complex superconductors, considering an extra internal DOF which can acquire two flavours. We find that there are many more possibilities to construct a nonunitary order parameter, some of which do not break any symmetry besides $U(1)$. Focusing on d-electron systems, we find that multiple point group symmetries allow for nonunitary order parameters in the trivial irreducible representation. Concerning the spectra, we find that particle-hole and particle-particle band crossings are protected for unitary order parameters with $\hat{F}_c(\bk) = 0$, but that the particle-hole crossings are lifted as soon as $\hat{F}_c(\bk) \neq 0$. Considering nonunitary order parameters, the crossings are again protected for $\hat{F}_c(\bk) = 0$, but both particle-particle and particle-hole crossings are lifted as soon as $\hat{F}_c(\bk) \neq 0$. In conclusion, the particle-hole crossings away from the Fermi energy are lifted as soon as $\hat{F}_c(\bk) \neq 0$, and particle-particle crossings require both $\hat{F}_c(\bk) \neq 0$ and a nonunitary order parameter. The later type of gap opening can be used as an indicator of nonunitary superconductivity.

Nonunitary order parameters have been recently discussed in multiple contexts, motivated by materials characterization and theoretical investigations. It was proposed that one of the signatures of nonunitary order parameters is the opening of gaps away from the Fermi surface in Dirac material \cite{Lado:2019}. Previous work has associated these gap openings to a measure of odd-frequency pairing correlations in Ising \cite{Tang:2021} and multi-band superconductors \cite{Komendova:2015}. The discussion in this manuscript comes as a reference to clarify how these spectral signatures proposed for the identification of nonunitary superconducting states and odd-frequency correlations are all tied to the same underlying concept of superconducting fitness.

An interesting perspective of this work is the general association of nonunitary order parameters with TRSB. In complex superconductors with multiple realizations for the order parameter within a single irreducible representation, a TRSB order parameter is possible within the trivial irreducible representation. The discussion of the stabilization this type of order parameter is left for future investigation.


\ack The author thanks Daniel F. Agterberg, Philip M. R. Brydon, and Carsten Timm for useful discussions. The author also acknowledges the financial support of the Swiss National Science Foundation through the Ambizione Grant No.~186043.


\section*{Appendix A}\label{sec:appendixA}

The point group $D_{2h}$ consists of the following operations: $E$, the identity; $C_{2n}$, two-fold rotations along the axis $n=x,y,z$; $i$ inversion; and $\sigma_{m}$, mirror operations along the planes $m=xy,xz,yz$. As can be seen from the character table (Tab. \ref{tab:character}), the irreducible representations can be uniquely identified by considering the character of the operations $C_{2x},C_{2y},C_{2z}$, and $i$.

\begin{table}[h]
\begin{center}
    \begin{tabular}{| c | c | c | c | c | c | c | c | c |}
    \hline
    irrep &   $E$ & $C_{2z}$ & $C_{2x}$ & $C_{2y}$ & $i$& $\sigma_{xy}$ & $\sigma_{xz}$ & $\sigma_{yz}$\\ \hline
    $A_{1g}$ & 1 & 1 & 1 & 1 & 1 & 1 & 1 & 1 \\ \hline
        $B_{1g}$ & 1 & 1 & -1 & -1 & 1 & 1 & -1 & -1 \\ \hline
            $B_{2g}$ & 1 & -1 &- 1 & 1 & 1 & -1 & 1 & -1 \\ \hline
                $B_{3g}$ & 1 & -1 & 1 & -1 & 1 & -1 &- 1 & 1 \\ \hline
    $A_{1u}$ & 1 & 1 & 1 & 1 & -1 & -1 & -1 & -1 \\ \hline
        $B_{1u}$ & 1 & 1 & -1 & -1 & -1 & -1 & 1 & 1 \\ \hline
            $B_{2u}$ & 1 & -1 &- 1 & 1 &- 1 & 1 & -1 & 1 \\ \hline
                $B_{3u}$ & 1 & -1 & 1 & -1 & -1 & 1 &1 & -1 \\ \hline
    \end{tabular}
        \end{center}
    \caption{  \label{Tab:D2h} Character table of the $D_{2h}$ point group.}
    \label{tab:character}
\end{table}

When acted upon by these operations, the coordinates transform as:
\begin{eqnarray}\label{Eq:coord}
C_{2x}&:& \{x,y,z\} \rightarrow \{x,-y,-z\},\\
C_{2y}&:& \{x,y,z\} \rightarrow \{-x,y,-z\},\\
C_{2z}&:& \{x,y,z\} \rightarrow \{-x,-y,z\},\\
i&:& \{x,y,z\} \rightarrow \{-x,-y,-z\}.
\end{eqnarray}

If we choose $\{d_{xz},d_{yz}\}$ as basis orbitals, these orbitals transform as:
\begin{eqnarray}
C_{2x}&:& \{d_{xz}, d_{yz}\} \rightarrow \{-d_{xz}, d_{yz}\}\Rightarrow -\hat{\tau}_3,\\
C_{2y}&:& \{d_{xz}, d_{yz}\} \rightarrow \{d_{xz}, -d_{yz}\} \Rightarrow \hat{\tau}_3,\\
C_{2z}&:& \{d_{xz}, d_{yz}\} \rightarrow \{-d_{xz}, -d_{yz}\}\Rightarrow -\hat{\tau}_0,\\
i&:& \{d_{xz}, d_{yz}\} \rightarrow \{d_{xz}, d_{yz}\}\Rightarrow \hat{\tau}_0,
\end{eqnarray}
such that we can associate a given $\hat{\tau}_i$ matrix in orbital space to each transformation.

Concerning the spin DOF, these transformations act as follows:
\begin{eqnarray}
C_{2x}&:& e^{-i \hat{\sigma}_x \pi/2} = -i  \hat{\sigma}_x,\\
C_{2y}&:& e^{-i \hat{\sigma}_y \pi/2} = -i  \hat{\sigma}_y,\\
C_{2z}&:& e^{-i \hat{\sigma}_z \pi/2} = -i  \hat{\sigma}_z,\\
i&:&\hat{\sigma}_0.
\end{eqnarray}

The complete matrix form of the point group operations above are then:
\begin{eqnarray}
C_{2x}&:& i \hat{\tau}_3 \otimes \hat{\sigma}_x,\\
C_{2y}&:& i \hat{\tau}_3 \otimes \hat{\sigma}_y,\\
C_{2z}&:& i \hat{\tau}_0 \otimes \hat{\sigma}_z,\\
i&:& \hat{\tau}_0 \otimes \hat{\sigma}_0.
\end{eqnarray}

The order parameter transforms under any unitary operation as $\hat{\Delta}(\bk) \rightarrow \hat{\Delta}'({\bk'}) = U \hat{\Delta}({\bk'}) U^T$, where the superscript $T$ indicates the transpose and $\bk' = U\bk$ is the rotated $\bk$ vector. Note that since the inversion operation acts trivially on the matrix structure of the order parameters, their parity in the EP scenario is directly determined by the parity of the accompanying momentum dependent function $d_{ab}(\bk)$, which is pre-determined by fermionic antisymmetry. Applying this prescription for the classification of order parameters, we find Table \ref{tab:irreps} in the main text.

\section*{Appendix B}\label{sec:appendixB}

Below are tables proving the classification of s-wave ($\bk$-independent) order parameters for two-orbital models with distinct pairs of d-electrons considering different point groups.

\begin{table}[H]
\begin{center}
    \begin{tabular}{| c | c | c | c | c |}
    \hline
[a,b]  & $\{d_{xz},d_{yz}\}$ & $\{d_{xz},d_{xy}\}$ & $\{d_{yz},d_{xy}\}$ &  $\{d_{x^2-y^2},d_{z^2}\}$ \\ \hline
[0,0]  & $\bold{A_g}$     & $\bold{A_g}$      & $\bold{A_g}$       & $\bold{A_g}$  \\ \hline
[1,0]  & $B_{1g}$ & $B_{3g}$ & $B_{2g}$   & $\bold{A_g}$ \\ \hline
[3,0]  & $\bold{A_g}$     & $\bold{A_g}$      & $\bold{A_g}$      & $\bold{A_g}$ \\ \hline
[2,1]  & $B_{2g}$ & $\bold{A_g}$     & $B_{1g}$  & $B_{3g}$ \\ \hline
[2,2]  & $B_{3g}$ & $B_{1g}$ & $\bold{A_g}$      & $B_{2g}$  \\ \hline
[2,3]  & $\bold{A_g}$     & $B_{2g}$ & $B_{3g}$     &  $B_{1g}$\\ \hline
    \end{tabular}
    \caption{$D_{2h}$ point group. The results for $\{d_{xz},d_{yz}\}$ also apply to $\{d_{xy},d_{x^2-y^2/z^2}\}$, the results for $\{d_{xz},d_{xy}\}$ also apply for $\{d_{yz},d_{x^2-y^2/z^2}\}$, and the results for  $\{d_{yz},d_{xy}\}$ also apply to $\{d_{xz},d_{x^2-y^2/z^2}\}$.}
    \label{tab:d2h}
\end{center}
\end{table}

\begin{table}[H]
\begin{center}
    \begin{tabular}{| c | c | c | c | c | c |}
    \hline
[a,b]  & $\{d_{xz},d_{yz}\}$ & $\{d_{xz},d_{xy}\}$ & $\{d_{yz},d_{xy}\}$  & $\{d_{x^2-y^2},d_{z^2}\}$ \\ \hline
[0,0]  & $\bold{A_1}$ & $\bold{A_1}$ &  $\bold{A_1}$ & $\bold{A_1}$ \\ \hline
[1,0]  & $A_2$ & $B_2$ &  $B_1$ &  $\bold{A_1}$\\ \hline
[3,0]  & $\bold{A_1}$ & $\bold{A_1}$ &  $\bold{A_1}$  & $\bold{A_1}$\\ \hline
[2,1]  & $B_1$ & $\bold{A_1}$  & $A_2$  & $B_2$\\ \hline
[2,2]  & $B_2$ & $A_2$  & $\bold{A_1}$ & $B_1$\\ \hline
[2,3]  & $\bold{A_1}$ & $B_1$ & $B_2$   & $A_2$\\ \hline
    \end{tabular}
    \caption{$C_{2v}$ point group. The results for $\{d_{xz},d_{yz}\}$ also apply to $\{d_{xy},d_{x^2-y^2/z^2}\}$, the results for $\{d_{xz},d_{xy}\}$ also apply for $\{d_{yz},d_{x^2-y^2/z^2}\}$, and the results for  $\{d_{yz},d_{xy}\}$ also apply to $\{d_{xz},d_{x^2-y^2/z^2}\}$.}
    \label{tab:c2v}
\end{center}
\end{table}

\clearpage
\begin{table}[H]
\begin{center}
    \begin{tabular}{| c | c | c | c | c |}
    \hline
[a,b]  & $\{d_{xz},d_{yz}\}$ & $\{d_{xy},d_{x^2-y^2}\}$ & $\{d_{xy},d_{z^2}\}$ & $\{d_{x^2-y^2},d_{z^2}\}$ \\ \hline
[0,0]  & $\bold{A_{1g}}$ & $\bold{A_{1g}}$ & $\bold{A_{1g}}$ & $\bold{A_{1g}}$ \\ \hline
[1,0]  & $B_{2g}$ & $A_{2g}$ & $B_{2g}$ &  $B_{1g}$\\ \hline
[3,0]  & $B_{1g}$ & $\bold{A_{1g}}$ & $\bold{A_{1g}}$ &  $\bold{A_{1g}}$\\ \hline
\{[2,1],[2,2]\}  & $E_g$ & $E_g$ & $E_g$  & $E_g$  \\ \hline
[2,3]  & $\bold{A_{1g}}$ & $\bold{A_{1g}}$ & $B_{1g}$ & $B_{2g}$\\ \hline
    \end{tabular}
    \caption{$D_{4h}$ point group.}
    \label{tab:d4h}
\end{center}
\end{table}

\begin{table}[H]
\begin{center}
    \begin{tabular}{| c | c | c | c | c |}
    \hline
[a,b]  & $\{d_{xz},d_{yz}\}$ & $\{d_{xy},d_{x^2-y^2}\}$ & $\{d_{xy},d_{z^2}\}$ & $\{d_{x^2-y^2},d_{z^2}\}$ \\ \hline
[0,0]  & $\bold{A_{1}}$ & $\bold{A_{1}}$ & $\bold{A_{1}}$ & $\bold{A_{1}}$ \\ \hline
[1,0]  & $B_{2}$ & $A_{2}$ & $B_{2}$ &  $B_{1}$\\ \hline
[3,0]  & $B_{1}$ & $\bold{A_{1}}$ & $\bold{A_{1}}$ &  $\bold{A_{1}}$\\ \hline
\{[2,1],[2,2]\}  & $E$ & $E$ & $E$  & $E$  \\ \hline
[2,3]  & $\bold{A_{1}}$ & $\bold{A_{1}}$ & $B_{1}$ & $B_{2}$\\ \hline
    \end{tabular}
    \caption{For the $D_{2d}$ point group.}
    \label{tab:d2d}
\end{center}
\end{table}

\begin{table}[H]
\begin{center}
    \begin{tabular}{| c | c |}
    \hline
[a,b]  & $\{d_{xz},d_{yz}\}$  \\ \hline
[0,0]  & $\bold{A_{1g}}$  \\ \hline
\{[1,0],[3,0]\} & $E_{2g}$  \\ \hline
\{[2,1],[2,2]\}  & $E_{1g}$   \\ \hline
[2,3]  & $\bold{A_{1g}}$  \\ \hline
    \end{tabular}
    \caption{For the $D_{6h}$ point group. Same results apply to $\{d_{xy},d_{x^2-y^2}\}$.}
    \label{tab:d6h}
\end{center}
\end{table}

\begin{table}[H]
\begin{center}
    \begin{tabular}{| c | c |}
    \hline
[a,b]  & $\{d_{xz},d_{yz}\}$ \\ \hline
[0,0]  & $\bold{A_{1}}$  \\ \hline
\{[1,0],[3,0]\} & $E_{2}$ \\ \hline
\{[2,1],[2,2]\}  & $E_1$   \\ \hline
[2,3]  & $\bold{A_{1}}$ \\ \hline
    \end{tabular}
    \caption{For the $C_{6v}$ point group. The same results apply to $\{d_{xy},d_{x^2-y^2}\}$.}
    \label{tab:c6v}
\end{center}
\end{table}


\section*{References}
\bibliographystyle{unsrt}
\bibliography{Nonunitary_arxiv}{}

\end{document}